\documentclass[lettersize,journal]{IEEEtran}
\usepackage{amsmath,amsfonts}
\usepackage{algorithmic}
\usepackage{algorithm}
\usepackage{array}
\usepackage[caption=false,font=normalsize,labelfont=sf,textfont=sf]{subfig}
\usepackage{textcomp}
\usepackage{stfloats}
\usepackage{url}
\usepackage{verbatim}
\usepackage{graphicx}
\usepackage{cite}

\begin{document}

\title{User Association and Transmit Beamforming for Rotatable Antenna-Enabled LAE Networks}
\author{Bowen Yao,~\IEEEmembership{Graduate Student Member,~IEEE}, Chao Zhang,~\IEEEmembership{Graduate Student Member,~IEEE}, 
	
and Ying-Chang Liang,~\IEEEmembership{Fellow,~IEEE}
\thanks{B. Yao and C. Zhang are with the National Key Laboratory of Wireless Communications, University of Electronic Science and Technology of China, Chengdu 611731, China (e-mail: bowen$\_$yao@std.uestc.edu.cn, zhang$\_$chao@std.uestc.edu.cn).}
\thanks{Y.-C. Liang is with the Institute for Fundamental and Frontier Sciences, University of Electronic Science and Technology of China, Chengdu 611731, China (e-mail: liangyc@ieee.org).}}


\maketitle

\begin{abstract}
This paper investigates rotatable antenna (RA)-enabled low-altitude economy (LAE) networks, where multiple base stations (BSs) are deployed to serve both terrestrial and aerial users. Unlike conventional fixed orientation antennas, RAs can adjust their boresights to serve users at diverse azimuth and elevation angles, thereby providing three-dimensional (3D) coverage. To improve the performance of LAE networks, we formulate a network sum-rate maximization problem that jointly optimizes user association, transmit beamforming, and RA orientations. However, in such LAE networks, aerial users are often visible to multiple BSs through line-of-sight (LoS) links, making user association and interference management particularly challenging. Moreover, user association and RA orientations are closely coupled: user association affects the RA pointing directions, while the RA orientations in turn reshape the channel conditions for user association. To solve this problem, we propose an alternating optimization (AO) algorithm that integrates weighted minimum mean-square error (WMMSE) for transmit beamforming, projected gradient ascent updates for user association, and Frank–Wolfe updates for RA orientations. Furthermore, we develop a discrete candidate-scanning scheme that replaces continuous orientation optimization with a finite search to reduce computational complexity. We also propose an antenna block-based scheme to reduce the number of orientation variables. Simulation results show that the proposed joint design achieves a higher sum-rate than the fixed antenna orientation and nearest BS association benchmarks. Moreover, the candidate-scanning and block-based schemes achieve approximately \textbf{89.9\%} and \textbf{94.3\%} of the performance with high transmit power, respectively, while reducing the complexity of orientation updates.
\end{abstract}


\section{Introduction}

The low-altitude economy (LAE) is an emerging economic paradigm that leverages unmanned aerial vehicles (UAVs) and electric vertical takeoff and landing vehicles (eVTOLs) to support diverse applications, such as logistics, environmental monitoring, and emergency response \cite{jiang20236g,Wang2025toward,he2026ubiquitous,he2025satellite}. As a promising market, LAE has drawn increasing attention worldwide, and there has been intensive research in designing LAE networks from both academia and industry recently, including, e.g., three-dimensional (3D) coverage \cite{azari2017coexistence,chen2023dedicating,chowdhury2021uptilted}, integrated sensing, communications and target identification \cite{zeng2026fmcw,jiang2025integrated,zeng2026in}, and UAV-enabled mobile edge computing \cite{mcenroe,jointyu}.

To provide 3D coverage for both terrestrial and aerial users in LAE, reusing existing terrestrial infrastructure is considered a cost-effective approach for LAE networks \cite{mishra2020survey,fotouhi2019survey,zeng2019cellular}. However, conventional base station (BS) antennas are typically down-tilted to prioritize terrestrial users, leading aerial users being served only by the sidelobes with poor antenna gains~\cite{yang2019optimal,geraci2018supporting}. To address the challenge, several studies have explored adjusting the antenna tilt angles to balance terrestrial and aerial coverage. In \cite{azari2017coexistence}, the authors derived the downlink coverage probability for both user types and showed that appropriately adjusting the down-tilted angles of the antennas can improve the performance of the networks. Moreover, the authors in \cite{chen2023dedicating} proposed converting a fraction of existing down-tilted BSs into up-tilted BSs, and examined how the fraction of up-tilted BSs, the up-tilted angle, and the beamwidth of BS antennas affect the coverage performance. Furthermore, a new cellular architecture was proposed in \cite{chowdhury2021uptilted}, in which each BS employs both down-tilted and up-tilted antennas to serve terrestrial users and UAVs, respectively. The up-tilted angles were optimized to maximize the minimum signal-to interference ratio among UAVs, while inter-cell interference coordination was employed to suppress interference from the down-tilted antennas.

Despite these advances, existing antenna tilt adjustment schemes still employ a fixed configuration, where the antennas at each BS share a common boresight direction. Consequently, even when the antennas are up-tilted, this common boresight cannot be aligned simultaneously with all distributed users across diverse azimuth and elevation angles. This misalignment inevitably reduces the antenna gain toward the intended users. To overcome this limitation, rotatable antennas (RAs) have recently emerged as a promising solution, enabling the boresight of each antenna to be independently adjusted without changing its position \cite{zheng2026rotatable2,zheng2026opportunities}. This additional degree of freedom facilitates considerable performance gains in terms of precise beam alignment, broader coverage, and interference management. Driven by these potential benefits, there has been several research to explore theoretical modeling and transmission design for RA systems. Specifically, a general modeling framework for RA-enabled wireless communications was established in \cite{zheng2026rotatable}, accompanied by an efficient channel estimation method developed in \cite{xiong2025estimation}. Moreover, a polarization-aware RA architecture that accounts for the polarization orientation of the radiated field was investigated and then applied to symbiotic radio systems \cite{zhang2026polarization,zhang2026polarizationsr}. For spectrum sharing, \cite{peng2026rotatable} jointly optimized RA orientations and transmit beamforming to maximize the weighted sum-rate of interference channels. For LAE networks, \cite{li2026rotatablelae} presented RA-aided multi-BS and multi-UAV cooperative coverage architectures and discussed cellular access and interference management, motivating coordinated transmission design.

However, the above RA studies generally assume a predetermined user association. This assumption is restrictive in LAE networks, where aerial users often maintain line-of-sight (LoS) links with multiple BSs. Such multi-BS visibility provides several candidate serving BSs but also exposes aerial users to severe interference. Moreover, the high mobility of aerial users leads to frequent handoffs between BSs, making user association particularly important to select an appropriate target BS. Prior studies have already investigated user association in LAE networks. In \cite{mei2019cellular}, the authors jointly optimized uplink cell association and power allocation to maximize the weighted sum-rate of a UAV and coexisting ground users. In \cite{hou2021joint}, UAV association, receive beamforming, and UAV heights were jointly optimized to maximize the minimum UAV rate while satisfying the rate requirements of terrestrial users. However, these studies do not account for the coupling between user association and RA orientations. Specifically, user association affects the RA orientations, while the RA directions alter the channel conditions, which in turn affect serving BS selection. This coupling calls for joint optimization of user association and RA orientations in LAE networks.

Motivated by the above, we investigate an RA-enabled LAE network in which multiple BSs serve both terrestrial and aerial users. We maximize the network sum-rate by jointly optimizing user association, transmit beamforming, and the orientations of RAs. The resulting design determines the serving BS of each user and coordinates transmit beamforming with RA orientations according to the 3D locations of terrestrial and aerial users. This coordination enhances desired signals, mitigates multi-user interference, and reduces antenna gain loss caused by boresight misalignment. Furthermore, for practical deployment, we develop two low-complexity algorithms for RA orientation optimization. The main contributions are summarized as follows:
\begin{itemize}
    \item We propose an RA-enabled LAE network in which multiple BSs adjust antenna boresights to simultaneously serve terrestrial and aerial users to support 3D coverage. For the proposed network, we formulate a sum-rate maximization problem that jointly optimizes user association, transmit beamforming, and RA orientations.
    \item To solve this problem, we first reformulate it as an equivalent weighted minimum mean-square error (WMMSE) problem and develop an alternating optimization (AO) algorithm. Specifically, user association, transmit beamforming, and the orientations of RAs are updated using projected gradient ascent, the WMMSE algorithm, and the Frank-Wolfe algorithm, respectively.
    \item To reduce the computational complexity for practical deployment, we propose two low-complexity algorithms for RA orientation optimization. Specifically, the discrete candidate-scanning algorithm restricts each RA boresight to a finite set of feasible directions determined by user locations, while the antenna block-based algorithm partitions the RAs into several groups, with all RAs within each group sharing a common boresight direction.
    \item Simulation results demonstrate that jointly optimizing user association, transmit beamforming, and the RA orientations substantially improves the sum-rate over fixed nearest BS association and fixed orientation benchmarks. Moreover, the low-complexity algorithms achieve approximately \text{89.9\%} and \text{94.3\%} of the performance while reducing the computational complexity of RA orientation updates.
\end{itemize}

The rest of this paper is organized as follows. Section II outlines the system model, covering the RA configuration, channel, and signal models. Section III formulates the network sum-rate maximization problem. Section IV presents the AO algorithm and its variable updates, followed by analyses of convergence and computational complexity. Section V proposes two low-complexity algorithms. Section VI shows simulation results, and Section VII concludes the study.

\textit{Notation:} Uppercase and lowercase bold letters denote matrices and vectors, respectively. Superscripts $(\cdot)^{T}$, $(\cdot)^{H}$, $(\cdot)^{*}$, and $(\cdot)^{-1}$ denote transpose, Hermitian transpose, complex conjugation, and inversion, respectively. The sets of $a\times b$ real and complex matrices are denoted by $\mathbb R^{a\times b}$ and $\mathbb C^{a\times b}$, respectively, while $\mathcal O(\cdot)$ denotes the computational complexity order. For a scalar $x$, $|x|$ and $\Re\{x\}$ denote its magnitude and real part, respectively. For a vector $\mathbf x$, $\|\mathbf x\|_2$ denotes its $\ell_2$-norm. Moreover, $[x]_+=\max\{x,0\}$, $\lfloor x\rfloor$ denotes the floor of $x$, and $[x]_n$ denotes the remainder of $x$ modulo $n$. The operators $\mathbb E\{\cdot\}$, $\nabla_{\mathbf x}$, and $\Pi_{\Omega}(\cdot)$ denote statistical expectation, the gradient with respect to $\mathbf x$, and Euclidean projection onto the set $\Omega$, respectively. The symbols $\mathbf I_N$, $\mathbf 0$, and $\mathbf 1$ denote the $N\times N$ identity matrix, an all-zero vector or matrix, and an all-one vector, respectively. Finally, superscripts $(t)$ and $\star$ indicate the iteration index and an optimal solution, respectively.

\section{System Model}
\label{system model}
\begin{figure}[!t]
	\centering
	\includegraphics[width=\columnwidth]{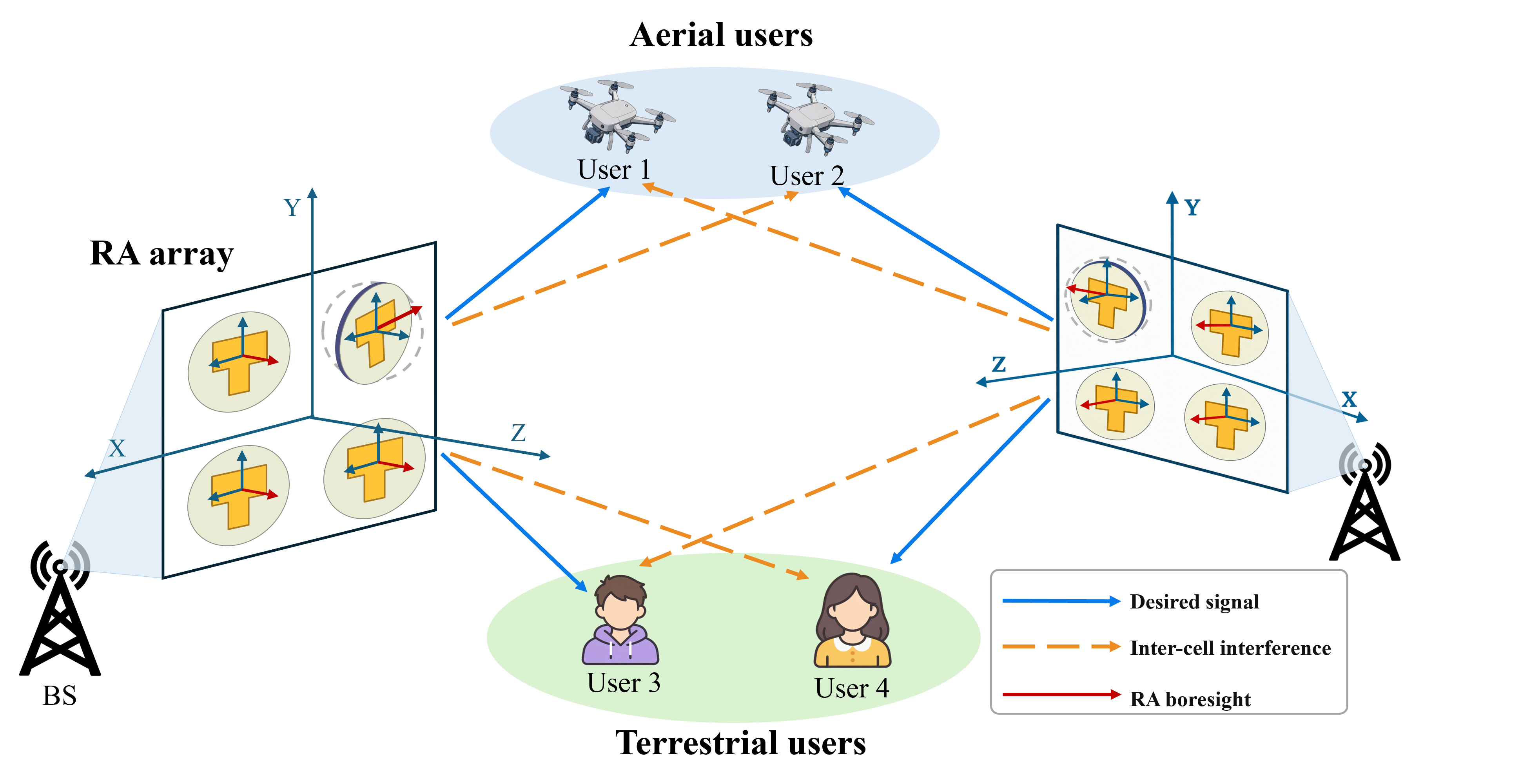}
	\caption{The RA-enabled LAE network.}
	\label{fig_1}
\end{figure}

As illustrated in Fig. \ref{fig_1}, we consider an RA-enabled LAE network, which consists of $ B $ BSs and $K$ users, denoted by the sets $ \mathcal{B} = \{1,\ldots,B\} $ and $ \mathcal{K} = \{1,\ldots,K\}$, respectively. In this network, the users include both conventional terrestrial users and low-altitude aerial users. Each BS is equipped with RAs to serve both types of users. Without loss of generality, we assume that each user is equipped with a single isotropic antenna and each BS employs a uniform planar array (UPA) with $M = M_x \times M_y$ antennas, where $M_x$ and $M_y$ denote the number of antennas along the $x$- and $y$- axes, respectively. 

\subsection{RA Configuration}
To support 3D coverage for LAE networks, we deploy RAs at multiple BSs where each RA has an independently adjustable boresight direction. Specifically, the boresight direction of the $m$-th RA at BS $b$ is characterized by a pointing vector
\begin{equation}
	\mathbf f_{b,m}
	=
	\begin{bmatrix}
		f_{b,m}^{x},\, f_{b,m}^{y},\, f_{b,m}^{z}
	\end{bmatrix}^{T}
	\in\mathbb R^{3\times1},
	\label{eq:ra_pointing_vector}
\end{equation}
where $f_{b,m}^{x}$, $f_{b,m}^{y}$, and $f_{b,m}^{z}$ denote the components along the $x$-, $y$-, and $z$-axes, respectively. Let $\theta_{b,m}\in[0,\frac{\pi}{2}]$ denote the zenith angle (i.e. the angle between the boresight direction and the positive $z$-axis), and let $\varphi_{b,m}\in[0,2\pi)$ denote the azimuth angle (i.e. the angle between the projection onto the $x$-$y$ plane and the positive $x$-axis). The pointing vector can then be expressed as
\begin{equation}
	\mathbf f_{b,m}
	=
	\begin{bmatrix}
		\sin\theta_{b,m}\cos\varphi_{b,m}\\
		\sin\theta_{b,m}\sin\varphi_{b,m}\\
		\cos\theta_{b,m}
	\end{bmatrix}.
	\label{eq:ra_angular_parameterization}
\end{equation}
To account for the limited rotation range, the zenith angle of each RA is subject to 
\begin{equation}
	0\leq\theta_{b,m}\leq\theta_{\max},
	\label{eq:ra_rotation_angle_range}
\end{equation}
where $\theta_{\max}\in[0,\frac{\pi}{2}]$ denotes the maximum zenith angle. Accordingly, the matrix of the RA orientations at BS $b$ is given by
$ \mathbf{F}_b = \begin{bmatrix} \mathbf{f}_{b,1}, \mathbf{f}_{b,2}, \ldots, \mathbf{f}_{b,M} \end{bmatrix} \in\mathbb{R}^{3\times M}$. The collection of orientation matrices across all BSs is then denoted by 
\begin{equation}
\mathbf{F} = \left\{ \mathbf{F}_b \mid b\in\mathcal{B} \right\}.
\end{equation}

To highlight the advantage of adjusting RA orientations, we adopt a directional radiation pattern based on \cite{zheng2026rotatable}, given by
\begin{equation}
	G\left(\epsilon\right)
	=
	\begin{cases}
		G_{\max}\bigl(\cos(\epsilon)\bigr)^{2p},
		& \epsilon \in \left[0,\dfrac{\pi}{2}\right),
		\\[2mm]
		0,
		& \text{otherwise},
	\end{cases}
	\label{eq:ra_gain_piecewise}
\end{equation}
where $\epsilon$ denotes the angular misalignment between the boresight direction and the signal propagation direction, $p\geq0$ is the antenna directivity factor, and $ G_{\max}=2(2p+1) $ is the maximum directional gain.

\subsection{Channel Model}
\label{subsec:channel_model}

In this subsection, we present a channel model for the considered RA-enabled LAE networks.

Let $\mathbf{p}_b \in \mathbb{R}^{3\times 1}$ denote the center position of the UPA at BS $b$ in the global Cartesian coordinates, and let $\mathbf{q}_k \in \mathbb{R}^{3\times 1}$ denote the position of user $k$. Each RA is indexed by $(m_x,m_y)$, where $ m_x \in \{1,\ldots,M_x\}$ and $ m_y \in \{1,\ldots,M_y\} $.
For notational convenience, the two-dimensional indices of RAs can be presented by a single index:
\begin{equation}
	m = m_x + (m_y-1)M_x,\quad \forall m \in \mathcal M,
\end{equation}
where $\mathcal M$ is defined as $\{1,\ldots,M\}$.
The position of the $m$-th RA at BS $b$ is then given by
\begin{equation}    
	\mathbf{t}_{b,m} = \mathbf{p}_b + \frac{\lambda}{2}
	\begin{bmatrix}        
		\left[m-1\right]_{M_x} - \frac{M_x-1}{2} \\[1.5mm]        
		\left\lfloor\frac{m-1}{M_x}\right\rfloor - \frac{M_y-1}{2} \\[1.5mm]        
		0    
	\end{bmatrix},
	\quad    
	\forall b\in\mathcal{B},\ m\in\mathcal{M},
	\label{eq:global_element_position}
\end{equation}
where $\lambda$ is the wavelength. The distance between the $m$-th RA at BS $b$ and user $k$ is denoted as $ r_{b,k,m}=\left\|\mathbf{q}_k-\mathbf{t}_{b,m}\right\|_2 $, and the corresponding unit propagation direction is $ \mathbf{u}_{b,k,m} = \frac{\mathbf{q}_k-\mathbf{t}_{b,m}}{ r_{b,k,m}}$. Based on the RA radiation pattern in Eq. \eqref{eq:ra_gain_piecewise}, the directional antenna gain of the $m$-th antenna at BS $b$ toward user $k$ is expressed as
\begin{equation}
	G_{b,k,m}
	\left(
	\mathbf{f}_{b,m}
	\right)
	=
	G_{\max}
	\left[
	\mathbf{f}_{b,m}^{T}
	\mathbf{u}_{b,k,m}
	\right]_+^{2p}.
	\label{eq:element_directional_gain}
\end{equation}
Thus, based on the Friis transmission model \cite{balanis1982antenna}, the channel between the $m$-th antenna at BS $b$ and user $k$ is given by
\begin{equation}
	h_{b,k,m}\!\left(\mathbf{f}_{b,m}\right)
	\! = \!
	\sqrt{ \! \beta_0 r_{b,k,m}^{-2}
	\! G_{b,k,m}\!\left(\mathbf{f}_{b,m}\right)}
	\exp\!\left(
	-\mathrm{j}\frac{2\pi}{\lambda}
	\boldsymbol{\ell}_{b,k}^{T}\mathbf{r}_{b,m}
	\right)\!,
	\label{eq:element_channel_exact}
\end{equation}
where $\beta_0=(\frac{\lambda}{4\pi})^2$ denotes the free space channel power gain at the reference distance, $\boldsymbol{\ell}_{b,k}=\frac{\mathbf q_k-\mathbf p_b}{\|\mathbf q_k-\mathbf p_b\|_2}$ denotes the unit direction vector from BS $b$ to user $k$, while $\mathbf r_{b,m}=\mathbf t_{b,m}-\mathbf p_b$ represents the relative position of the $m$-th RA with respect to the center position of BS $b$. Thus, the channel vector from BS $b$ to user $k$ can be written as
\begin{equation}
	\mathbf{h}_{b,k}
	\left(
	\mathbf{F}_b
	\right)
	=
	\begin{bmatrix}
		h_{b,k,1}
		\left(
		\mathbf{f}_{b,1}
		\right),
		\ldots,
		h_{b,k,M}
		\left(
		\mathbf{f}_{b,M}
		\right)
	\end{bmatrix}^{T}
	\in\mathbb{C}^{M\times 1}.
	\label{eq:exact_channel_vector}
\end{equation}


\subsection{Signal Model}
\label{subsec:signal_association_model}

In this subsection, we develop a signal model combined with user association where each user is associated with only one BS. Specifically, let $a_{b,k}\in\{0,1\}$ denote the user association variable for all $b\in\mathcal{B}$ and $k\in\mathcal{K}$, where $a_{b,k}=1$ if and only if user $k$ is associated with BS $b$. Since each user is served by only one BS, the association variables satisfy
\begin{equation}
	\sum_{b\in\mathcal{B}}a_{b,k}=1,
	\qquad
	\forall k\in\mathcal{K}.
	\label{eq:unique_association}
\end{equation}
We denote the set of all association variables as $\mathbf{A}= \left\{ a_{b,k} \mid b\in\mathcal{B},\;k\in\mathcal{K} \right\}$.

For the transmission, let $s_{b,k}$ denote the transmit data intended for user $k$ from BS $b$ satisfying $ \mathbb{E}\left\{|s_{b,k}|^2\right\}=1 $, and let $ \mathbf{v}_{b,k}\in\mathbb{C}^{M\times 1} $ denote the corresponding transmit beamforming vector. Then the signal from BS $b$ can be expressed as
\begin{equation}
	\mathbf{x}_b
	=
	\sum_{k\in\mathcal{K}}
	\mathbf{v}_{b,k}s_{b,k}.
	\label{eq:bs_transmitted_signal}
\end{equation}
It is worth noting that $\mathbf{v}_{b,k}$ is set to zero when user $k$ is not associated with BS $b$, i.e., $\mathbf{v}_{b,k}=\mathbf{0}$ if $a_{b,k}=0$. Accordingly, the received signal at user $k$ is given by
\begin{equation}
	y_k
	=
	\sum_{b \in\mathcal{B} }\mathbf{h}_{b,k}^{H}\mathbf{x}_{b}+
	z_k,
	\label{eq:received_signal_at_user_k}
\end{equation}
where $z_k\sim\mathcal{CN}(0,\sigma_k^2)$ represents the additive white Gaussian noise (AWGN) at user $k$. When user $k$ is served by BS $b$, its corresponding signal-to-interference-plus-noise ratio (SINR) is given by
\begin{equation}
	\gamma_{b,k} =
	\frac{
		\left|
		\mathbf{h}_{b,k}^{H}
		\mathbf{v}_{b,k}
		\right|^2
	}{
		\displaystyle
		\sum_{n \neq k}
		\left|\mathbf{h}_{b,k}^{H} \mathbf{v}_{b,n}\right|^2
		+
		\sum_{\substack{j\in\mathcal{B} \\ j \neq b}}
		\sum_{\ell \in\mathcal{K}}
		\left|
		\mathbf{h}_{j,k}^{H}
		\mathbf{v}_{j,\ell}
		\right|^2
		+
		\sigma_k^2
	}.
	\label{eq:candidate_sinr}
\end{equation}
Then the achievable rate is expressed as
\begin{equation}
 r_{b,k} =	\log_2 \left( 1+ \gamma_{b,k} \right).
\end{equation}

\section{Problem Formulation}
\label{sec:problem_formulation}

Based on the above system model, the user association affects the optimal RA orientations, since different serving BSs result in different propagation directions. However, it is not straightforward to determine the RA orientations jointly with the user association. Therefore, we consider a network sum-rate maximization problem which jointly optimizes user association $\mathbf{A}$, transmit beamforming $\mathbf{V}=\left\{\mathbf{v}_{b,k}\mid b\in\mathcal{B},\,k\in\mathcal{K}\right\}$ and the RA orientations $\mathbf{F}$. The resulting optimization problem is formulated as
\begin{subequations}
	\label{prob:joint_optimization}
	\begin{align}
		\text{\textbf{P}1:}\quad
		\max_{\mathbf{A},\mathbf{V},\mathbf{F}}
		\quad&
		\sum_{k\in\mathcal{K}}
		\sum_{b\in\mathcal{B}}
		a_{b,k}r_{b,k}
		\label{prob:joint_objective}
		\\
		\text{s.t.}\quad&
		\sum_{k\in\mathcal{K}}
		\left\lVert\mathbf{v}_{b,k}\right\rVert_2^2
		\leq P_b,
		\quad
		\forall b\in\mathcal{B},
		\label{prob:bs_power_constraint}
		\\
		&
		a_{b,k}\in\{0,1\},
		\quad
		\forall b\in\mathcal{B},\;
		k\in\mathcal{K},
		\label{prob:binary_association_constraint}
		\\
		&
		\sum_{b\in\mathcal{B}}a_{b,k}=1,
		\quad
		\forall k\in\mathcal{K},
		\label{prob:unique_association_constraint}
		\\
		&
		\left\lVert\mathbf{f}_{b,m}\right\rVert_2=1,
		\quad
		\forall b\in\mathcal{B},\;
		m\in\mathcal{M},
		\label{prob:orientation_unit_norm}
		\\
		&
		\cos\theta_{\mathrm{max}}
		\leq
		\mathbf{n}_b^{T}\mathbf{f}_{b,m}
		\leq 1,
		\quad
		\forall b\in\mathcal{B},\;
		m\in\mathcal{M}.
		\label{prob:orientation_rotation_constraint}
	\end{align}
\end{subequations}

The objective of problem \textbf{P}1 is to maximize the network sum-rate, defined as the sum of $r_{b,k}$ weighted by their corresponding association variables. Constraint \eqref{prob:bs_power_constraint} imposes an individual transmit power budget at each BS, where $P_b$ denotes the maximum transmit power of BS $b$. Moreover, constraints \eqref{prob:binary_association_constraint} and \eqref{prob:unique_association_constraint} jointly define the feasible set for user association. Specifically, the binary variable $a_{b,k}=1$ indicates that user $k$ is associated with BS $b$, and $a_{b,k}=0$ otherwise, while constraint \eqref{prob:unique_association_constraint} ensures that each user is served by only one BS. Constraint \eqref{prob:orientation_unit_norm} ensures that the orientation vector $\mathbf{f}_{b,m}$ of each RA is normalized. Constraint \eqref{prob:orientation_rotation_constraint} limits the maximum rotation angle $\theta_{\max}$ of each RA boresight $\mathbf f_{b,m}$ from the initial direction $\mathbf n_b$. Given that $\mathbf n_b^{T}\mathbf f_{b,m}=\cos\theta_{b,m}$,  this is equivalent to $0\leq\theta_{b,m}\leq\theta_{\max}$.

It is worth noting that problem \textbf{P}1 does not explicitly impose the coupling constraint $\mathbf v_{b,k}=0$ when $a_{b,k}=0$ because omitting this constraint does not affect the optimal objective value. To see this, suppose that there exists an optimal solution satisfying $a^{\star}_{b,k}=0$ but $\mathbf v^{\star}_{b,k}\neq\mathbf 0$. Setting $\mathbf v^{\star}_{b,k} = \mathbf 0$ does not affect the corresponding sum-rate, since $a^{\star}_{b,k}r_{b,k} = 0$. Meanwhile, setting $\mathbf v^{\star}_{b,k}=0$ reduces the transmit power consumption at BS $b$ and cannot increase the interference experienced by any served user. Therefore, the system sum-rate cannot decrease. Hence, there always exists an optimal solution satisfying $\mathbf v_{b,k}=0$ whenever $a_{b,k}=0$ for all $b\in\mathcal B$ and $k\in\mathcal K$.


\section{Proposed Solution}
\label{sec:proposed solution}

Problem \textbf{P}1 is challenging to solve directly, as it requires the joint optimization of the user association, transmit beamforming, and RA orientations. Moreover, the effective channels depend on the RA orientations, making the problem non-convex.

To address these challenges, we first reformulate the sum-rate objective using the WMMSE equivalence by introducing the auxiliary variables $\mathbf U$ and $\mathbf W$. Then, we develop an AO algorithm that sequentially updates the auxiliary variables $\mathbf U$ and
$\mathbf W$, the user association $\mathbf A$, the transmit beamforming $\mathbf V$, and the RA orientations $\mathbf F$, while keeping the remaining variables fixed.

\subsection{Problem Reformulation}
\label{subsec:problem_reformulation}

In this subsection, we reformulate the non-convex sum-rate objective into an equivalent WMMSE form to facilitate the joint optimization \cite{shi2011iteratively}. Specifically, we introduce the auxiliary variable $u_{b,k}\in\mathbb C$, which denotes the scalar receive equalizer associated with the link from BS \(b\) to user \(k\). The estimate of $s_{b,k}$ is then given by $\hat{s}_{b,k}=u_{b,k}^{*}y_k$. The corresponding mean square error (MSE) is
\begin{equation}
	e_{b,k}
	=
	\mathbb E \left[
	\left|\hat{s}_{b,k}-s_{b,k}\right|^2
	\right].
	\label{eq:mse_definition}
\end{equation}
Substituting the received signal given by Eq. \eqref{eq:received_signal_at_user_k} into Eq.~\eqref{eq:mse_definition} yields
\begin{equation}
	e_{b,k}
	=|u_{b,k}|^2T_k
	-2\Re\!\left\{
	u_{b,k}^{*}\mathbf h_{b,k}^{H}\mathbf v_{b,k}
	\right\}+1,
	\label{eq:mse_expansion}
\end{equation}
where
$T_k=\sum_{\ell\in\mathcal B}\sum_{j\in\mathcal K}
\left|\mathbf h_{\ell,k}^{H}\mathbf v_{\ell,j}\right|^2+\sigma_k^2$
denotes the total received power at user $k$.

For fixed transmit beamforming $\mathbf V$, minimizing $e_{b,k}$ with respect to $u_{b,k}$ leads to the minimum mean square error (MMSE) equalizer
\begin{equation}
	u_{b,k}^{\star}
	=\frac{\mathbf h_{b,k}^{H}\mathbf v_{b,k}}{T_k}.
	\label{eq:mmse_equalizer}
\end{equation}
By substituting Eq. \eqref{eq:mmse_equalizer} into Eq. \eqref{eq:mse_expansion}, the corresponding MMSE is then obtained as
\begin{equation}
	e_{b,k}^{\rm mmse}
	=1-
	\frac{\left|\mathbf h_{b,k}^{H}\mathbf v_{b,k}\right|^2}{T_k}
	=\frac{1}{1+\gamma_{b,k}}.
	\label{eq:mmse_sinr_relation}
\end{equation}
Therefore, the achievable rate from BS $b$ to user $k$ can be equivalently expressed as $r_{b,k}=-\log_2 e_{b,k}^{\rm mmse}$. By introducing a positive weight $w_{b,k}>0$, the following identity holds for any $e_{b,k}>0$
\begin{equation}
	-\ln e_{b,k}
	=\max_{w_{b,k}>0}
	\left(\ln w_{b,k}-w_{b,k}e_{b,k}+1\right).
	\label{eq:wmmse_identity}
\end{equation}
Solving the maximization problem in Eq.~\eqref{eq:wmmse_identity} with respect to $w_{b,k}$ yields the optimal weight
\begin{equation}
	w_{b,k}^{\star}=e_{b,k}^{-1}.
	\label{eq:wmmse_weight}
\end{equation}
By combining Eq.~\eqref{eq:mmse_sinr_relation} with Eq.~\eqref{eq:wmmse_identity}, the achievable rate from BS $b$ to user $k$ can be equivalently expressed as
\begin{equation}
	r_{b,k}
	=\max_{u_{b,k},\,w_{b,k}>0}
	\frac{\ln w_{b,k}-w_{b,k}e_{b,k}+1}{\ln 2}.
	\label{eq:rate_wmmse_equivalence}
\end{equation}
Consequently, the original sum-rate maximization problem \textbf P1 can be equivalently reformulated as the following problem
\begin{align}
	\text{\textbf{P}2:}\quad
	\max_{\mathbf U,\mathbf W,\mathbf A,\mathbf V,\mathbf F}\quad&
	\frac{1}{\ln 2}
	\sum_{k\in\mathcal K}
	\sum_{b\in\mathcal B}
	a_{b,k}
	\left(
	\ln w_{b,k}
	-
	w_{b,k}e_{b,k}
	+
	1
	\right)
	\label{prob:wmmse_reformulation}
	\\
	\mathrm{s.t.}\quad&
	\eqref{prob:bs_power_constraint}\text{--}
	\eqref{prob:orientation_rotation_constraint},
	\notag
\end{align}
where $\mathbf U=\{u_{b,k}\mid b\in\mathcal B,\,k\in\mathcal K\}$ and $\mathbf W=\{w_{b,k}\mid b\in\mathcal B,\,k\in\mathcal K\}$. For any fixed $\mathbf A$, $\mathbf V$, and $\mathbf F$, the optimal variables $u_{b,k}^{\star}$ and $w_{b,k}^{\star}$ are given by Eq. \eqref{eq:mmse_equalizer} and Eq.~\eqref{eq:wmmse_weight}, respectively.

\subsection{Optimization for User Association}
\label{subsec:user_association_optimization}
In this subsection, we optimize the user association $\mathbf A$ with fixed $\mathbf U$, $\mathbf W$, $\mathbf V$, and $\mathbf F$. Since the transmit power constraint \eqref{prob:bs_power_constraint} and the RA orientation constraints \eqref{prob:orientation_unit_norm}--\eqref{prob:orientation_rotation_constraint} are independent of $\mathbf A$, they can be omitted from the user association subproblem. The resulting subproblem is formulated as
\begin{align}
	\text{\textbf{P}3:}\quad
	\max_{\mathbf A}\quad&
	\frac{1}{\ln 2}
	\sum_{k\in\mathcal K}
	\sum_{b\in\mathcal B}
	a_{b,k}
	\left(
	\ln w_{b,k}
	-
	w_{b,k}e_{b,k}
	+
	1
	\right)
	\label{prob:association_subproblem}
	\\
	\mathrm{s.t.}\quad&
	\eqref{prob:binary_association_constraint},\;
	\eqref{prob:unique_association_constraint}.
	\notag
\end{align}

Problem \textbf P3 is non-convex because of the binary association constraint \eqref{prob:binary_association_constraint}. To facilitate this optimization, we relax $a_{b,k}$ from the binary set $\{0,1\}$ to $[0,1]$. Together with constraint~\eqref{prob:unique_association_constraint}, the relaxed feasible set is defined as 
\begin{equation}
	\Omega
	=
	\left\{
		\mathbf a \in \mathbb R^{B\times 1}
		\mid
		\mathbf a \ge \mathbf 0,
		\mathbf 1^T \mathbf a = 1
		\right\},
	\label{eq:probability_simplex}
\end{equation}
where $\mathbf a_k =[a_{1,k},\ldots,a_{B,k}]^T \in \Omega $ for all $k \in \mathcal K$.
 
Accordingly, the problem \textbf P3 can be formulated as
\begin{subequations}
	\label{prob:relaxed_association}
	\begin{align}
		\text{\textbf{P}3.1:}\quad
		\max_{\{\mathbf a_k\}_{k\in\mathcal K}}
		\quad&
		\sum_{k\in\mathcal K}
		\mathbf r_k^{T}\mathbf a_k
		\label{prob:relaxed_association_obj}
		\\
		\mathrm{s.t.}\quad&
		\mathbf a_k\in\Omega,
		\qquad \forall k\in\mathcal K,
		\label{prob:relaxed_association_constraint}
	\end{align}
\end{subequations}
where $\mathbf r_k=[r_{1,k},\ldots,r_{B,k}]^{T}$.

It is obvious that problem \textbf P3.1 has an optimal solution where user $k$ assigns its entire association weight to a single BS $b_k \in\operatorname*{arg\,max}_{b\in\mathcal B}a_{b,k}$. Hence, there exists an optimal solution in which $a_{b,k}$ is binary for all $b \in \mathcal B$ and $k \in \mathcal K$. Consequently, the relaxed problems \textbf P3.1 is equivalent to problem \textbf P3 and attains the same optimal objective value. However, solving problem \textbf P3.1 directly to optimality may yield a low-quality solution because the AO algorithm is highly sensitive to the initial point. To solve it, we therefore perform a projected gradient ascent step mentioned in \cite{sanjabi2014optimal} at each iteration as
\begin{equation}
	\mathbf a_k^{(t+1)}
	=
	\Pi_{\Omega}
	\left(
	\mathbf a_k^{(t)}
	+
	\lambda^{(t)}
	\mathbf r_k^{(t)}
	\right),
	\qquad \forall k\in\mathcal K,
	\label{eq:association_projection_update}
\end{equation}
where $\lambda^{(t)}>0$ is the stepsize and $\Pi_{\Omega}(\cdot)$ denotes the Euclidean projection onto the set $\Omega$. Details of the projection procedure are provided in Appendix~\ref{app:association_projection_derivation}.


After the AO iterations, a binary recovery step is performed to enforce the constraint \eqref{prob:binary_association_constraint}. Specifically, for each user $k$, we select $b_k\in\operatorname*{arg\,max}_{b\in\mathcal B}a_{b,k}$, and then recover the binary association as
\begin{equation}
	a_{b,k}^{\rm bin}
	=
	\begin{cases}
		1, & b=b_k,
		\\
		0, & b\neq b_k.
	\end{cases}
	\label{eq:association_binary_recovery}
\end{equation}

\subsection{Optimization for Transmit Beamforming}
\label{subsec:beamforming_optimization}

In this subsection, we optimize the transmit beamforming $\mathbf V$ with fixed $\mathbf U$, $\mathbf W$, $\mathbf A$, and $\mathbf F$. Since the terms $\ln w_{b,k}+1$ and the positive factor $1/\ln 2$ are independent of $\mathbf V$, problem \textbf P2 can be reduced to the following subproblem
\begin{align}
	\text{\textbf{P}4:}\quad
	\min_{\mathbf V}\quad&
	\sum_{k\in\mathcal K}
	\sum_{b\in\mathcal B}
	a_{b,k}w_{b,k}e_{b,k}
	\label{prob:beamforming_mse}
	\\
	\mathrm{s.t.}\quad&
	\eqref{prob:bs_power_constraint}.
	\notag
\end{align}

Substituting Eq.~\eqref{eq:mse_expansion} into the objective function in Eq.~\eqref{prob:beamforming_mse} and omitting the terms independent of $\mathbf V$, problem \textbf P4 can be decomposed into $B$ independent subproblems. For BS $b$, the corresponding subproblem is given by
\begin{equation}
	\label{prob:per_bs_beamforming1}
	\begin{aligned}
		\text{\textbf{P}4.1:}\quad
		\min_{\{\mathbf v_{b,k}\}_{k\in\mathcal K}}\quad&
		\sum_{k\in\mathcal K}
		\mathbf v_{b,k}^{H}
		\mathbf C_b
		\mathbf v_{b,k}
		\\[-0.2ex]
		&{}-
		2\sum_{k\in\mathcal K}
		a_{b,k}w_{b,k}
		\Re\!\left\{
		u_{b,k}^{*}
		\mathbf h_{b,k}^{H}
		\mathbf v_{b,k}
		\right\}
		\\[0.3ex]
		\mathrm{s.t.}\quad&
		\eqref{prob:bs_power_constraint},
	\end{aligned}
\end{equation}
where $\mathbf C_b$ is denoted as $\sum_{\!\ell \in \mathcal{B}} \sum_{\!j \in \mathcal{K}} a_{\ell,j}w_{\ell,j} \left| u_{\ell,j} \right|^2 \mathbf h_{b,j} \mathbf h_{b,j}^{H}$.

Let $\mu_b\geq0$ denote the Lagrange multiplier associated with the power constraint~\eqref{prob:bs_power_constraint}. Then, the Lagrangian of problem \textbf P4.1 is given by
\begin{equation}
	\label{eq:beamforming_lagrangian}
	\begin{aligned}
		\mathcal L_b
		={}&
		\sum_{k\in\mathcal K}
		\mathbf v_{b,k}^{H}
		\mathbf C_b
		\mathbf v_{b,k}
		\\[-0.2ex]
		&{}-
		2\sum_{k\in\mathcal K}
		a_{b,k}w_{b,k}
		\Re\!\left\{
		u_{b,k}^{*}
		\mathbf h_{b,k}^{H}
		\mathbf v_{b,k}
		\right\}
		\\[-0.2ex]
		&{}+
		\mu_b
		\left(
		\sum_{k\in\mathcal K}
		\left\lVert
		\mathbf v_{b,k}
		\right\rVert_2^2
		-P_b
		\right).
	\end{aligned}
\end{equation}
Taking the Wirtinger derivative with respect to $\mathbf v_{b,k}^{*}$ gives
\begin{equation}
	\frac{\partial\mathcal L_b}
	{\partial\mathbf v_{b,k}^{*}}
	=
	\left(
	\mathbf C_b+\mu_b\mathbf I_M
	\right)
	\mathbf v_{b,k}
	-
	a_{b,k}w_{b,k}u_{b,k}\mathbf h_{b,k}.
	\label{eq:beamforming_lagrangian_gradient}
\end{equation}
According to the Karush–Kuhn–Tucker (KKT) stationarity condition, the optimal transmit beamforming vector is given by
\begin{equation}
	\mathbf v_{b,k}^{\star}
	=
	a_{b,k}w_{b,k}u_{b,k}
	\left(
	\mathbf C_b+\mu_b\mathbf I_M
	\right)^{-1}
	\mathbf h_{b,k},
	\forall b\in\mathcal B,\! k\in\mathcal K.
	\label{eq:optimal_beamforming_vector}
\end{equation}
The Lagrange multiplier $\mu_b$ is determined independently for each BS via bisection to satisfy the power constraint~\eqref{prob:bs_power_constraint}.


\subsection{Optimization for the Orientations of RAs}
\label{subsec:orientation_optimization}

In this subsection, we optimize the RA orientations $\mathbf F$ with $\mathbf U$, $\mathbf W$, $\mathbf A$, and $\mathbf V$ fixed. The resulting subproblem is formulated as
\begin{align}
	\text{\textbf{P}5:}\quad
	\max_{\mathbf F}\quad&
	\mathcal G_F(\mathbf F)
	\label{prob:orientation_subproblem}\\
	\mathrm{s.t.}\quad&
	\mathbf f_{b,m}\in\mathcal F_b,
	\quad
	\forall b\in\mathcal B,\;m\in\mathcal M.
	\nonumber
\end{align}
Here, $\mathcal G_F$ denotes the objective function of problem~\textbf{P}2 in Eq.~\eqref{prob:wmmse_reformulation}, and $\mathbf F$ is the collection of $\mathbf f_{b,m}$ for all $b \in \mathcal B$ and $m\in\mathcal M$. The feasible set $\mathcal F_b$ defined by constraints~\eqref{prob:orientation_unit_norm} and \eqref{prob:orientation_rotation_constraint} is
\begin{equation}
	\mathcal F_b
	=
	\left\{
	\mathbf x\in\mathbb R^{3\times1}
	\,\middle|\,
	\|\mathbf x\|_2=1,\;
	\cos\theta_{\max}
	\leq
	\mathbf n_b^T\mathbf x
	\leq1
	\right\}.
	\label{eq:orientation_feasible_cap}
\end{equation}

Problem~\textbf{P}5 is challenging due to the coupling between the RA orientations and the effective channels, together with the non-convex feasible set $\mathcal F_b$. To address these difficulties, we adopt a Frank-Wolfe method to optimize the RA orientations~\cite{FW,peng2026rotatable}. At each iteration, the procedure constructs an approximation of problem~\textbf{P}5 at the current feasible orientations, then solves the resulting subproblem, and updates the RA orientations using a stepsize selected by Armijo backtracking.

Specifically, at iteration $t$, we first compute the gradient of $\mathcal G_F(\mathbf F)$ with respect to the RA orientation $\mathbf f_{b,m}$, given by
\begin{equation}
	\mathbf g_{b,m}^{(t)}
	=
	\left.
	\nabla_{\mathbf f_{b,m}}
	\mathcal G_F(\mathbf F)
	\right|_{\mathbf F=\mathbf F^{(t)}},
	\label{eq:orientation_euclidean_gradient_definition}
\end{equation}
which is detailed in Appendix~\ref{app:orientation_gradient_derivation}. However, since constraint~\eqref{prob:orientation_unit_norm} restricts $\mathbf f_{b,m}^{(t)}$ to a unit sphere, we have to project the gradient $\mathbf g_{b,m}^{(t)}$ onto the tangent space as
\begin{equation}
	\mathbf q_{b,m}^{(t)}
	=
	\mathbf T_{b,m}^{(t)}
	\mathbf g_{b,m}^{(t)},
	\label{eq:orientation_tangent_gradient}
\end{equation}
where $\mathbf T_{b,m}^{(t)} = \mathbf I_3 - \mathbf f_{b,m}^{(t)} \left(\mathbf f_{b,m}^{(t)}\right)^{T}$ is the corresponding orthogonal projection matrix.

To simplify the optimization of the nonlinear objective $\mathcal G_F$, we then use the projected gradients to approximate $\mathcal G_F$ around the current orientations $\mathbf F^{(t)}$ as
\begin{align}
	&
	\mathcal G_F\!\left(\mathbf F^{(t)}\right)+
	\sum_{b\in\mathcal B}
	\sum_{m\in\mathcal M}
	\left(\mathbf q_{b,m}^{(t)}\right)^T
	\left(
	\mathbf f_{b,m}
	-
	\mathbf f_{b,m}^{(t)}
	\right)
	\nonumber\\
	&=
	\sum_{b\in\mathcal B}
	\sum_{m\in\mathcal M}
	\left(\mathbf q_{b,m}^{(t)}\right)^T
	\mathbf f_{b,m}
	\nonumber\\
	&\quad+
	\Biggl[
	\mathcal G_F\!\left(\mathbf F^{(t)}\right)
	-
	\sum_{b\in\mathcal B}
	\sum_{m\in\mathcal M}
	\left(\mathbf q_{b,m}^{(t)}\right)^T
	\mathbf f_{b,m}^{(t)}
	\Biggr].
	\nonumber
\end{align} 
Replacing the objective function $\mathcal G_F$ of problem~\textbf{P}5 with this approximation results in the following subproblem
\begin{align}
	\text{\textbf{P}5.1:}\quad
	\max_{\mathbf F}\quad&
	\sum_{b\in\mathcal B}
	\sum_{m\in\mathcal M}
	\left(\mathbf q_{b,m}^{(t)}\right)^T
	\mathbf f_{b,m}
	\nonumber\\[-0.5ex]
	&+
	\left[
	\mathcal G_F\!\left(\mathbf F^{(t)}\right)
	-
	\sum_{b\in\mathcal B}
	\sum_{m\in\mathcal M}
	\left(\mathbf q_{b,m}^{(t)}\right)^T
	\mathbf f_{b,m}^{(t)}
	\right]
	\label{prob:orientation_first_order_subproblem}\\
	\mathrm{s.t.}\quad&
	\mathbf f_{b,m}\in\mathcal F_b,
	\quad
	\forall b\in\mathcal B,\;m\in\mathcal M.
	\nonumber
\end{align}
After omitting the term independent of $\mathbf F$, problem~\textbf{P}5.1 can be decomposed into the following subproblem for each $\mathbf f_{b,m}$
\begin{align}
	\text{\textbf{P}5.2:}\quad
	\max_{\mathbf f_{b,m}}\quad&
	\left(\mathbf q_{b,m}^{(t)}\right)^T
	\mathbf f_{b,m}
	\label{prob:orientation_linear_oracle}\\
	\mathrm{s.t.}\quad&
	\mathbf f_{b,m}\in\mathcal F_b.
	\nonumber
\end{align}
We denote $\mathbf s_{b,m}^{(t)}$ as an optimal solution for problem~\textbf{P}5.2, given in Eq.~\eqref{eq:orientation_fw_oracle_closed_form}, where $\boldsymbol{\zeta}_b$ is any unit vector orthogonal to $\mathbf n_b$.
\begin{figure*}[!b]
	\noindent\rule{\textwidth}{0.4pt}
	\vspace{-1.5ex}
	\begin{equation}
		\resizebox{0.90\textwidth}{!}{$\displaystyle
			\mathbf s_{b,m}^{(t)}
			=
			\begin{cases}
				\frac{
					\mathbf q_{b,m}^{(t)}
				}{
					\|\mathbf q_{b,m}^{(t)}\|_2
				},
				&
				\|\mathbf q_{b,m}^{(t)}\|_2>0,\quad
				\left(\mathbf q_{b,m}^{(t)}\right)^{T}\mathbf n_b
				\geq
				\cos\theta_{\max}\|\mathbf q_{b,m}^{(t)}\|_2,
				\\[1.5ex]
				\cos\theta_{\max}\,\mathbf n_b
				+
				\sin\theta_{\max}
				\frac{
					\left(
					\mathbf I_3-\mathbf n_b\mathbf n_b^{T}
					\right)
					\mathbf q_{b,m}^{(t)}
				}{
					\left\|
					\left(
					\mathbf I_3-\mathbf n_b\mathbf n_b^{T}
					\right)
					\mathbf q_{b,m}^{(t)}
					\right\|_2
				},
				&
				\left(\mathbf q_{b,m}^{(t)}\right)^{T}\mathbf n_b
				<
				\cos\theta_{\max}\|\mathbf q_{b,m}^{(t)}\|_2,\quad
				\left\|
				\left(
				\mathbf I_3-\mathbf n_b\mathbf n_b^{T}
				\right)
				\mathbf q_{b,m}^{(t)}
				\right\|_2>0,
				\\[2ex]
				\cos\theta_{\max}\,\mathbf n_b
				+
				\sin\theta_{\max}\,\boldsymbol{\zeta}_b,
				&
				\left(\mathbf q_{b,m}^{(t)}\right)^{T}\mathbf n_b
				<
				\cos\theta_{\max}\|\mathbf q_{b,m}^{(t)}\|_2,\quad
				\left\|
				\left(
				\mathbf I_3-\mathbf n_b\mathbf n_b^{T}
				\right)
				\mathbf q_{b,m}^{(t)}
				\right\|_2=0,
				\\[1.5ex]
				\mathbf f_{b,m}^{(t)},
				&
				\|\mathbf q_{b,m}^{(t)}\|_2=0.
			\end{cases}
			$}
		\label{eq:orientation_fw_oracle_closed_form}
	\end{equation}
	\vspace{-1.5ex}
\end{figure*}
Consequently, the Frank-Wolfe search directions are given by
\begin{equation}
	\mathbf d_{b,m}^{(t)} = \mathbf s_{b,m}^{(t)} - \mathbf f_{b,m}^{(t)}\quad,
    \forall b\in\mathcal B,\; m\in\mathcal M.	
\end{equation}
Then, the antenna orientations can be updated as
\begin{equation}
	\mathbf f_{b,m}^{(t+1)}
	=
	\frac{
		\mathbf f_{b,m}^{(t)}
		+
		\rho^{(t)}\mathbf d_{b,m}^{(t)}
	}{
		\left\|
		\mathbf f_{b,m}^{(t)}
		+
		\rho^{(t)}\mathbf d_{b,m}^{(t)}
		\right\|_2
	},
	\quad
	\forall b\in\mathcal B,\;m\in\mathcal M,
	\label{eq:orientation_normalized}
\end{equation}
where $\rho^{(t)} \in (0,1]$ is the stepsize, selected by the following Armijo condition:
\begin{equation}
	\mathcal G_F(\mathbf F^{(t+1)})\geq \mathcal G_F(\mathbf F^{(t)})+c\rho^{(t)}\sigma^{(t)}.
	\label{armijo}
\end{equation}
\begin{algorithm}[t]
	\caption{The Frank-Wolfe Algorithm for Problem~\textbf{P}5}
	\label{alg:ra_orientation_fw}
	\begin{algorithmic}[1]
		\REQUIRE Fixed $\mathbf U$, $\mathbf W$, $\mathbf A$, and $\mathbf V$; initial feasible orientations $\mathbf F^{(0)}$; tolerance $\epsilon_{\mathrm{FW}}$; maximum iteration number $T_{\mathrm{FW}}$.		
		\ENSURE Optimized RA orientations $\mathbf F^{(t)}$.
		
		\STATE Set $t\gets0$ and compute $\mathcal G_F(\mathbf F^{(0)})$;
		
		\REPEAT
		\FOR{each $b\in\mathcal B$}
		\FOR{each $m\in\mathcal M$}
		\STATE Compute the gradient $\mathbf g_{b,m}^{(t)}$ using Eq.~\eqref{eq:orientation_euclidean_gradient_definition};
		\STATE Compute the projected gradient $\mathbf q_{b,m}^{(t)}$ using Eq.~\eqref{eq:orientation_tangent_gradient};
		\STATE Obtain the solution $\mathbf s_{b,m}^{(t)}$ of the subproblem~\textbf{P}5.2 using Eq.~\eqref{eq:orientation_fw_oracle_closed_form};
		\STATE Construct the search direction $\mathbf d_{b,m}^{(t)}=\mathbf s_{b,m}^{(t)}-\mathbf f_{b,m}^{(t)}$;
		\ENDFOR
		\ENDFOR
		\STATE Calculate $\sigma^{(t)} =\sum_{b\in\mathcal B}\sum_{m\in\mathcal M}\left(\mathbf q_{b,m}^{(t)}\right)^T\mathbf d_{b,m}^{(t)}$;
		\STATE Select the stepsize $\rho^{(t)}$ by Armijo backtracking, satisfying Eq.\eqref{armijo};
		\STATE Update the RA orientations $\mathbf F^{(t+1)}$using Eq.\eqref{eq:orientation_normalized};
		\STATE Compute $\mathcal G_F(\mathbf F^{(t+1)})$;
		\STATE Set $t\gets t+1$;
		\UNTIL{$t=T_{\mathrm{FW}}$ or 
			$\frac{|\mathcal G_F(\mathbf F^{(t)})-\mathcal G_F(\mathbf F^{(t-1)})|
			}{|\mathcal G_F(\mathbf F^{(t-1)}|}	\leq \epsilon_{\mathrm{FW}}	$}
		\STATE Output $\mathbf F^{(t)}$.	
	\end{algorithmic}
\end{algorithm}The parameter $c\in(0,1)$ in Eq.~\eqref{armijo} is an Armijo constant, while $\sigma^{(t)}$ denotes an aggregate improvement of the objective in problem~\textbf{P}5.2 for all $b\in\mathcal B$ and $m\in\mathcal M$. Specifically, it is defined as
\begin{equation}
	\begin{aligned}
		\sigma^{(t)}
		&=
		\sum_{b\in\mathcal B}
		\sum_{m\in\mathcal M}
		\left(\mathbf q_{b,m}^{(t)}\right)^T
		\left(
		\mathbf s_{b,m}^{(t)}
		-
		\mathbf f_{b,m}^{(t)}
		\right)\\
		&=
		\sum_{b\in\mathcal B}
		\sum_{m\in\mathcal M}
		\left(\mathbf q_{b,m}^{(t)}\right)^T
		\mathbf d_{b,m}^{(t)}.
	\end{aligned}
	\label{eq:orientation_fw_gap}
\end{equation}
Given that $\left(\mathbf q_{b,m}^{(t)}\right)^T\mathbf s_{b,m}^{(t)}\geq\left(\mathbf q_{b,m}^{(t)}\right)^T\mathbf f_{b,m}^{(t)}$ consistently holds, $\sigma^{(t)}\geq0$. Meanwhile, it is proved that there always exists a stepsize $\rho^{(t)}$ satisfying condition~\eqref{armijo} at each iteration~\cite{peng2026rotatable}. Therefore, $\mathcal G_F(\mathbf F^{(t+1)})\geq \mathcal G_F(\mathbf F^{(t)})$, guaranteeing the objective sequence $\{\mathcal G_F\}$ is monotonically non-decreasing. The procedure for optimizing the RA orientations is summarized in Algorithm~\ref{alg:ra_orientation_fw}. 

In summary, the overall algorithm to solve the problem \textbf P1 is summarized in Algorithm~\ref{alg:overall_ao}.
\begin{algorithm}[H]
	\caption{Proposed Algorithm for Problem \textbf P1}
	\label{alg:overall_ao}
	\begin{algorithmic}[1]
		\REQUIRE Initial feasible $\mathbf A^{(0)}$, $\mathbf V^{(0)}$, and $\mathbf F^{(0)}$; tolerance $\epsilon$; maximum iteration number $T_{\max}$.
		\ENSURE Optimized user association $\mathbf A^{(t)}$, transmit beamforming $\mathbf V^{(t)}$, and RA orientations $\mathbf F^{(t)}$.
		\STATE Set $t\gets0$ and evaluate the objective function $R^{(0)}$ of problem \textbf P1;
		\REPEAT
		\STATE Update $\mathbf U^{(t+1)}$ using Eq. \eqref{eq:mmse_equalizer};
		\STATE Update $\mathbf W^{(t+1)}$ using Eq. \eqref{eq:wmmse_weight};
		\STATE Update $\mathbf A^{(t+1)}$ using	Eq. \eqref{eq:association_projection_update};
		\STATE Update $\mathbf V^{(t+1)}$ using Eq. \eqref{eq:optimal_beamforming_vector};
		\STATE Update $\mathbf F^{(t+1)}$ using Algorithm~\ref{alg:ra_orientation_fw};
		\STATE Evaluate	$R^{(t+1)}$;
		\STATE Set $t\gets t+1$;
		\UNTIL{$t=T_{\max}$ or
			$
			\frac{
				|R^{(t)}-R^{(t-1)}|
			}{
				|R^{(t-1)}|
			}
			\leq \epsilon
			$}
		\STATE Output $\mathbf A^{(t)}$, $\mathbf V^{(t)}$, and $\mathbf F^{(t)}$.
	\end{algorithmic}
\end{algorithm}
\subsection{Convergence and Complexity Analysis}
\label{subsec:overall_ao}
\textit{1) Convergence Analysis:}
Let $R(\mathbf A,\mathbf V,\mathbf F)$ denote the objective function of problem \textbf{P1}. For updating the user association $\mathbf A$, Appendix~\ref{app:association_projection_derivation} establishes that the updating procedure does not decrease the objective value, as shown in Eq.~\eqref{eq:association_projection_ascent1}. In the subsequent beamforming update, $\mathbf V$ is optimized with $\mathbf A^{(t)}$ and $\mathbf F^{(t-1)}$ fixed. Since the subproblem \textbf{P}4 is convex, Eq.~\eqref{eq:optimal_beamforming_vector} provides an optimal solution. Because \(\mathbf V^{(t-1)}\) remains feasible with constraints unchanged, the update cannot increase the objective of problem~\textbf{P}4 and, equivalently, cannot decrease the objective $R$. We then update the RA orientations \(\mathbf F\) using the Frank-Wolfe method. According to the Armijo condition in Eq.\eqref{armijo}, the objective value of the subproblem \textbf{P}5 is monotonically non-decreasing. Consequently, we have $R^{(t-1)} \leq R^{(t)}$. Therefore, the sequence $\{R^{(t)}\}$ is monotonically non-decreasing.

Moreover, the bound $G_{b,k,m}\leq G_{\max}$, together with the transmit power constraint (13b), ensures that the objective value is bounded above. Hence, the monotonically non-decreasing sequence \(\{R^{(t)}\}\) is convergent.

\textit{2) Computational Complexity:}
We analyze the computational complexity in terms of the AO iterations. For the updates of $\mathbf U$ and $\mathbf W$, the optimal solutions are given by Eq.~\eqref{eq:mmse_equalizer} and Eq.~\eqref{eq:wmmse_weight}, respectively. The computational cost is dominated by evaluating $ T_k=\sum_{\ell\in\mathcal B}\sum_{j\in\mathcal K} \left|\mathbf h_{\ell,k}^{H}\mathbf v_{\ell,j}\right|^2+\sigma_k^2 $ for each user $k$, whose complexity is $\mathcal O(BKM)$. Aggregating this cost over all $K$ users yields $\mathcal O(BK^2M)$. For the update of $\mathbf A$, the main computational complexity comes from the projection in Eq.~\eqref{eq:association_projection_update}. As detailed in Appendix~\ref{app:association_projection_derivation}, the projection sorts the $B$ entries of the vector $\mathbf a_k^{(t)}+\lambda^{(t)}\mathbf r_k^{(t)}$ for each user $k$, which has a complexity of $\mathcal O(B\log B)$. Thus, updating $\mathbf A$ for all $K$ users has a complexity of $\mathcal O(KB\log B)$. For the update of $\mathbf V$ in Eq.~\eqref{eq:optimal_beamforming_vector}, the complexity of constructing $\mathbf C_b$ for all $b \in \mathcal B$ is $\mathcal O(BKM^2)$. Inverting each $M\times M$ matrix $\mathbf C_b+\mu_b\mathbf I_M$ has a complexity of $\mathcal O(M^3)$, totaling $\mathcal O(BM^3)$ for all BSs. Since the bisection search of $\mu_b$ generally takes only a few iterations, its complexity is omitted. Consequently, the overall complexity of updating $\mathbf V$ is $\mathcal O(BKM^2+BM^3)$. For the update of $\mathbf F$, the complexity is dominated by gradient evaluation in Eq.~\eqref{eq:appendix_orientation_gradient_expanded}. Each gradient evaluation requires a complexity of $\mathcal O\!\left(K(K+B)\right)$, leading to an overall complexity of $\mathcal O\!\left(BMK(K+B)\right)$. Let $t_{\mathrm{AO}}$ denote the number of AO iterations and  $t_{\mathrm{FW}}$ denote the number of Frank-Wolfe iterations per AO step. The total computational complexity is then given by
\begin{equation}
	\begin{aligned}
		\mathcal O\!\Bigg(
		t_{\mathrm{AO}}\Big[
		&\underbrace{BK^2M}_{\mathbf U,\mathbf W}
		+\underbrace{KB\log B}_{\mathbf A}\\
		&+\underbrace{BKM^2+BM^3}_{\mathbf V}
		+\underbrace{t_{\mathrm{FW}}BMK(K+B)}_{\mathbf F}
		\Big]\Bigg).
	\end{aligned}
	\label{eq:overall_computational_complexity}
\end{equation}

\section{Low-Complexity Joint Optimization Algorithms}
\label{sec:low_complexity_algorithms}

However, optimizing the RA orientations requires iterative updates, and the associated computational complexity increases with the number of RAs. In this section, we develop two low-complexity algorithms that simplify the optimization of RA orientations while leaving the updates of the remaining variables unchanged.

\subsection{Discrete Candidate-Scanning Algorithm}
\label{subsec:discrete_candidate_scanning}

In this subsection, we develop a discrete candidate-scanning-based algorithm that sequentially updates each RA orientation by selecting the direction from a finite candidate set that maximizes the objective. 

As indicated in Eq.~\eqref{eq:element_directional_gain}, the directional antenna gain of an RA is maximized when the boresight is aligned with the user direction. Therefore, the user directions provide a natural basis for constructing a finite set of candidate orientations. However, a user direction $\mathbf u_{b,k,m}$ may violate the rotation constraint~\eqref{prob:orientation_rotation_constraint}. To ensure feasibility, we define $\mathcal P_b(\mathbf x) = \operatorname*{arg\,max}_{\mathbf f\in\mathcal F_b} \mathbf f^{T}\mathbf x$ to project any user direction $\mathbf x$ onto the feasible region $\mathcal F_b$. Then the candidate set for the $m$-th RA at BS $b$ is constructed as
\begin{equation}
	\mathcal C_{b,m}^{(t)}
	=
	\left\{
	\mathbf f_{b,m}^{(t)}
	\right\}
	\cup
	\left\{
	\mathcal P_b(\mathbf u_{b,k,m})
	\ \middle|\ k\in\mathcal K
	\right\}.
	\label{eq:discrete_candidate_set}
\end{equation}
The current orientation $\mathbf f_{b,m}^{(t)}$ is included in $\mathcal C_{b,m}^{(t)}$ so that the objective value cannot decrease. With $\mathbf U$, $\mathbf W$, $\mathbf A$, and $\mathbf V$ fixed, the subproblem remains as formulated in problem~\textbf P5. Rather than evaluating the orientation gradient, the proposed method updates $\mathbf f_{b,m}$ by selecting a candidate orientation according to
\begin{equation}
	\mathbf f_{b,m}^{(t+1)}
	\in
	\operatorname*{arg\,max}_{
		\mathbf c\in\mathcal C_{b,m}^{(t)}
	}
	\mathcal G_F(\mathbf c).
	\label{eq:discrete_scan_update}
\end{equation}
Since $\mathbf f_{b,m}^{(t)}\in\mathcal C_{b,m}^{(t)}$, the update in Eq.~\eqref{eq:discrete_scan_update} satisfies
\begin{equation}
	\mathcal G_F
	\left(
	\mathbf F^{(t+1)}
	\right)
	\geq
	\mathcal G_F
	\left(
	\mathbf F^{(t)}
	\right).
	\label{eq:discrete_scan_monotonicity}
\end{equation}
The complete procedure is summarized in Algorithm~\ref{alg:discrete_candidate_scan}.

\begin{algorithm}[t]
	\caption{Discrete Candidate-Scanning Algorithm}
	\label{alg:discrete_candidate_scan}
	\begin{algorithmic}[1]
		\REQUIRE Initial feasible $\mathbf A^{(0)}$, $\mathbf V^{(0)}$,
		and $\mathbf F^{(0)}$; tolerance $\epsilon$;
		maximum iteration number $T_{\max}$.
		\ENSURE Optimized user association $\mathbf A^{(t)}$, transmit beamforming $\mathbf V^{(t)}$, and RA orientations $\mathbf F^{(t)}$.
		
		\STATE Set $t\gets 0$ and evaluate $R^{(0)}$;
		\REPEAT
		\STATE Update $\mathbf U^{(t+1)}$ using Eq.~\eqref{eq:mmse_equalizer};
		\STATE Update $\mathbf W^{(t+1)}$ using Eq.~\eqref{eq:wmmse_weight};
		\STATE Update $\mathbf A^{(t+1)}$ using Eq.~\eqref{eq:association_projection_update};
		\STATE Update $\mathbf V^{(t+1)}$ using Eq.~\eqref{eq:optimal_beamforming_vector};
		\FOR{each $b\in\mathcal B$}
		\FOR{each $m\in\mathcal M$}
		\STATE
		Construct $\mathcal C_{b,m}^{(t)}$ using Eq.~\eqref{eq:discrete_candidate_set};
		\STATE
	    Update $\mathbf f_{b,m}^{(t+1)}$ using Eq.~\eqref{eq:discrete_scan_update};
		\ENDFOR
		\ENDFOR
		\STATE Evaluate $R^{(t+1)}$;
		\STATE Set $t\gets t+1$;
		\UNTIL{$t=T_{\max}$ or
			$
			\frac{
				\left|R^{(t)}
				-R^{(t-1)}\right|
			}{
				\left|R^{(t-1)}\right|
			}
			\leq\epsilon$}
		
		\STATE Output $\mathbf A^{(t)}$, $\mathbf V^{(t)}$,
		and $\mathbf F^{(t)}$.
	\end{algorithmic}
\end{algorithm}

\subsection{Antenna Block-Based Algorithm}
\label{subsec:block_based_algorithm}

In this subsection, we develop a low-complexity antenna-block-based algorithm that reduces the number of orientation variables by partitioning the RAs at each BS into multiple blocks. All RAs within each block share a common boresight direction.

Let $\mathcal N_b=\{1,\ldots,N_b\}$ denote the block index set for BS $b$. We partition the RA index set $\mathcal M$ into $N_b$ blocks, satisfying
\begin{equation}
		\mathcal M =
		\bigcup_{n\in\mathcal N_b}\mathcal M_{b,n},
	\label{eq:block_partition}
\end{equation}
where $\mathcal M_{b,n}\subseteq\mathcal M$ denotes the set of RAs assigned to block $n\in\mathcal N_b$. For each $b\in\mathcal B$ and $n\in\mathcal N_b$, let $\mathbf f_{b,n}^{\rm blk}\in\mathcal F_b$ denote the shared boresight direction of block $n$ at BS $b$. Thus, every RA in $\mathcal M_{b,n}$ satisfies
\begin{equation}
	\mathbf f_{b,m}
	=
	\mathbf f_{b,n}^{\rm blk},
	\quad
	\forall m\in\mathcal M_{b,n}.
	\label{eq:block_shared_orientation}
\end{equation}
Accordingly, the subproblem can be written as
\begin{subequations}
	\label{prob:block_orientation_subproblem}
	\begin{align}
		\textup{\textbf P6:}\;\quad
		\max_{\mathbf F_{\rm blk}}\quad&
		\mathcal G_F\left(\mathbf F_{\rm blk}\right)
		\label{prob:block_orientation_objective}
		\\
		\mathrm{s.t.}\quad&
		\mathbf f_{b,n}^{\rm blk}\in\mathcal F_b,
		\quad
		\forall b\in\mathcal B,\ n\in\mathcal N_b,
		\label{prob:block_orientation_constraint}
	\end{align}
\end{subequations}
where $ \mathbf F_{\rm blk} = \left\{ \mathbf f_{b,n}^{\rm blk} \mid b\in\mathcal B,\ n\in\mathcal N_b \right\}$. Since all RAs within block $\mathcal M_{b,n}$ share the same boresight direction, we optimize only the shared variable $\mathbf f_{b,n}^{\rm blk}$ rather than the individual orientation of every RA in the block. Problem~\textbf{P}6 has the same constraint structure as problem~\textbf{P}5. Therefore, $\mathbf F_{\rm blk}$ is optimized following the same procedure described in Algorithm~\ref{alg:ra_orientation_fw}.

\section{Simulation Results}
\label{sec:simulation_results}

\begin{figure}[!t]
	\centering
	\includegraphics[width=0.5\columnwidth,
	keepaspectratio]{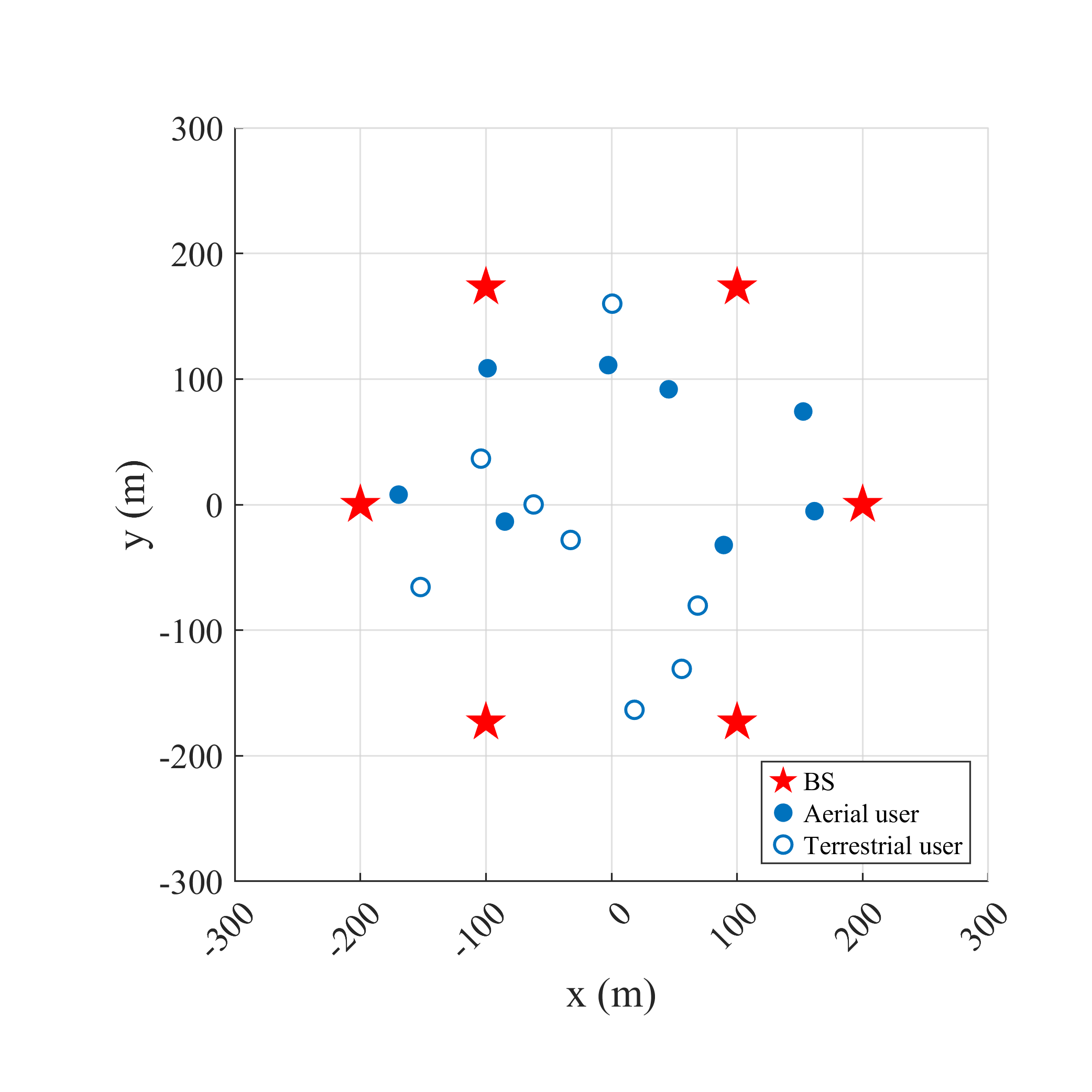}
	\caption{The simulation scenario.}
	\label{fig:simulation_setup}
\end{figure}

In this section, we evaluate the performance of the proposed algorithm. Unless otherwise specified, the simulation parameters are set as follows. As illustrated in Fig.~\ref{fig:simulation_setup}, six BSs are deployed in a regular hexagonal layout, each equipped with a $2\times2$ UPA oriented toward the network center. The distance between adjacent BSs and the BS height are set to \(200\) m and \(20\) m, respectively. The carrier wavelength is set to $\lambda=0.125$ m and the inter-element spacing is set to $d=0.5\lambda$. The network serves $K=16$ users, including eight terrestrial users at a height of $1.5$ m and eight aerial users whose heights are uniformly distributed over $[40,60]$ m. The transmit power per BS and the receiver noise power are set to $P_b=10$ dBm and $\sigma^2=-80$ dBm, respectively. The RA directivity factor is set to $p=2$ and the maximum rotation angle is set to $\theta_{\max}=\pi/3$. The key simulation parameters are summarized in Table~\ref{tab:simulation_parameters}.

\begin{table*}[!t]
	\centering
	\caption{Simulation Parameters}
	\label{tab:simulation_parameters}
	\renewcommand{\arraystretch}{1.15}
	\begin{tabular}{l c l c}
		\hline
		\textbf{Parameter} & \textbf{Value} &
		\textbf{Parameter} & \textbf{Value}\\
		\hline
		Number of BSs, $B$ & $6$ &
		Number of users, $K$ & $16$\\
		Distance between adjacent BSs & $200$ m &
		BS height & $20$ m\\
		Number of terrestrial users & $8$ &
		Terrestrial user height & $1.5$ m\\
		Number of aerial users & $8$ &
		Aerial user height & $[40,60]$ m\\
	    UPA size, $M_x\times M_y$ & $2\times2$ &	
	    Wavelength, $\lambda$ & $0.125$ m \\
		Inter-element spacing, $d$ & $0.5\lambda$ &
		Transmit power per BS, $P_b$ & $10$ dBm \\
		Receiver noise power, $\sigma^2$ & $-80$ dBm &
		Directivity factor, $p$ & $2$ \\
		Maximum rotation angle, $\theta_{\max}$ & $\pi/3$ & \\
		\hline
	\end{tabular}
\end{table*}

\subsection{Performance of the Proposed Algorithm}
\label{subsec:simulation_main_results}

In this subsection, we evaluate the performance gains of the proposed joint design in Algorithm~\ref{alg:overall_ao} compared with five benchmark schemes, which are defined as follows:
\begin{itemize}
	\item \textbf{Baseline 1 (fixed antenna orientations):}
	User association is optimized, and transmit beamforming is optimized using WMMSE, while the RA orientations are fixed.
	\item \textbf{Baseline 2 (nearest BS association):}
	Each user is associated with its nearest BS, whereas the RA orientations are optimized and transmit beamforming is optimized using WMMSE.
	\item \textbf{Baseline 3 (nearest BS association and fixed antenna orientations):}
    Each user is associated with its nearest BS, and the RA orientations are fixed. In this scheme, only transmit beamforming is optimized using WMMSE.
	\item \textbf{Baseline 4 (maximum ratio transmission, MRT):}
	User association and the RA orientations are optimized, while the MRT algorithm is used for beamforming.
	\item \textbf{Baseline 5 (zero forcing, ZF):}
	User association and the RA orientations are optimized, while the ZF algorithm is used for beamforming.
\end{itemize}
\begin{figure}[!t]
	\centering
	\includegraphics[
	width=0.75\columnwidth,
	keepaspectratio
	]{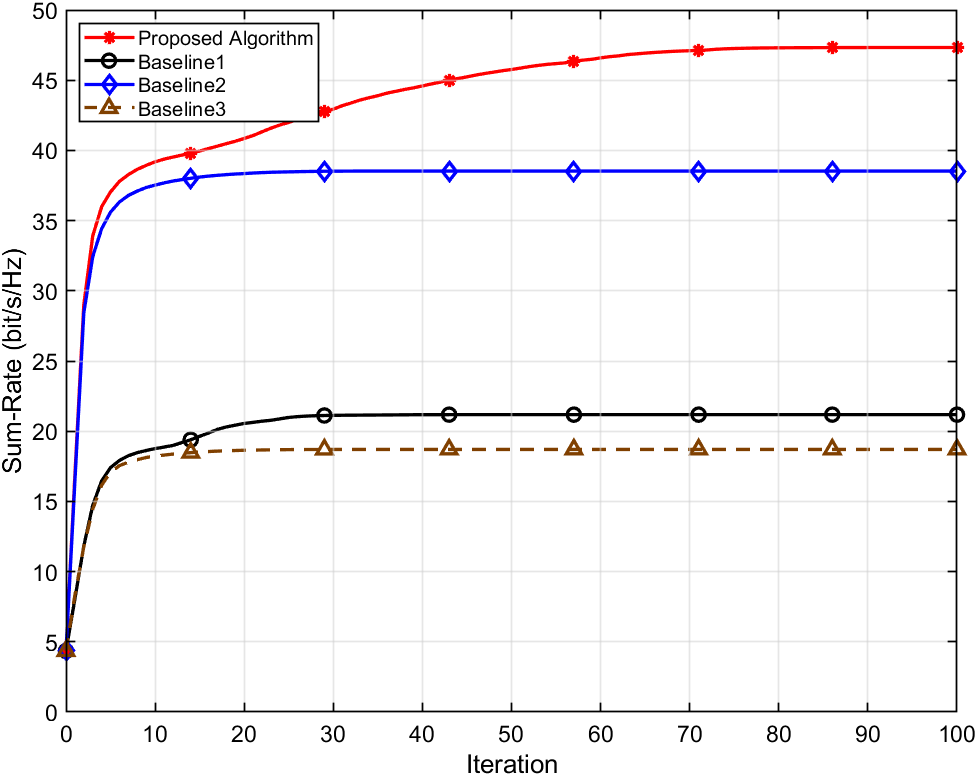}
	\caption{Convergence behavior.}
	\label{fig:simulation_convergence}
\end{figure}

Fig.~\ref{fig:simulation_convergence} illustrates the convergence behavior of the proposed algorithm alongside Baselines~1--3. It can be observed that all four curves increase monotonically over the iterations. At convergence, the proposed algorithm achieves the highest sum-rate compared with the three benchmarks, demonstrating its performance superiority. Moreover, Baseline~2 achieves a higher sum-rate than Baseline~1, indicating that the gain from optimizing the RA orientations outweighs that from optimizing user association.
\begin{figure}[!t]
	\centering
	\includegraphics[width=0.75\columnwidth]
	{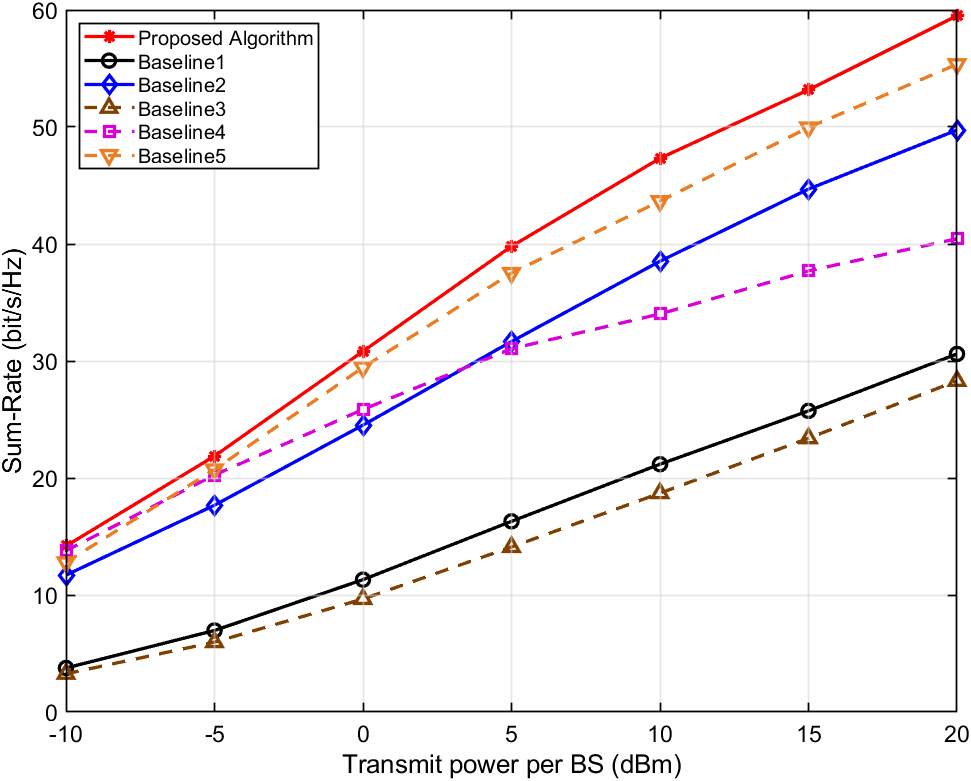}
	\caption{Sum-rate versus the transmit power per BS.}
	\label{fig:simulation_power}
\end{figure}

Fig.~\ref{fig:simulation_power} depicts the sum-rate versus transmit power per BS. All schemes exhibit a monotonic increase with transmit power, and the proposed algorithm consistently attains the highest sum-rate. At the lowest transmit power of $-10$ dBm, Baseline~4 with MRT performs nearly as well as the proposed algorithm with WMMSE and slightly outperforms Baseline~5 with ZF. This observation is consistent with a noise-limited scenario, in which enhancing the desired signal is more important than suppressing multi-user interference. As power increases, however, Baseline~5 surpasses Baseline~4 and thereafter remains the closest benchmark to the proposed algorithm, owing to the interference suppression capability of ZF.  Moreover, the gap between the proposed algorithm and the fixed orientation scheme Baseline~1 widens as transmit power increases, indicating that orientation optimization becomes increasingly beneficial in the interference-limited scenario, where RA can reshape the channel conditions to strengthen desired links and mitigate inter-user interference.
\begin{figure}[!t]
	\centering
	\includegraphics[width=0.75\columnwidth]
	{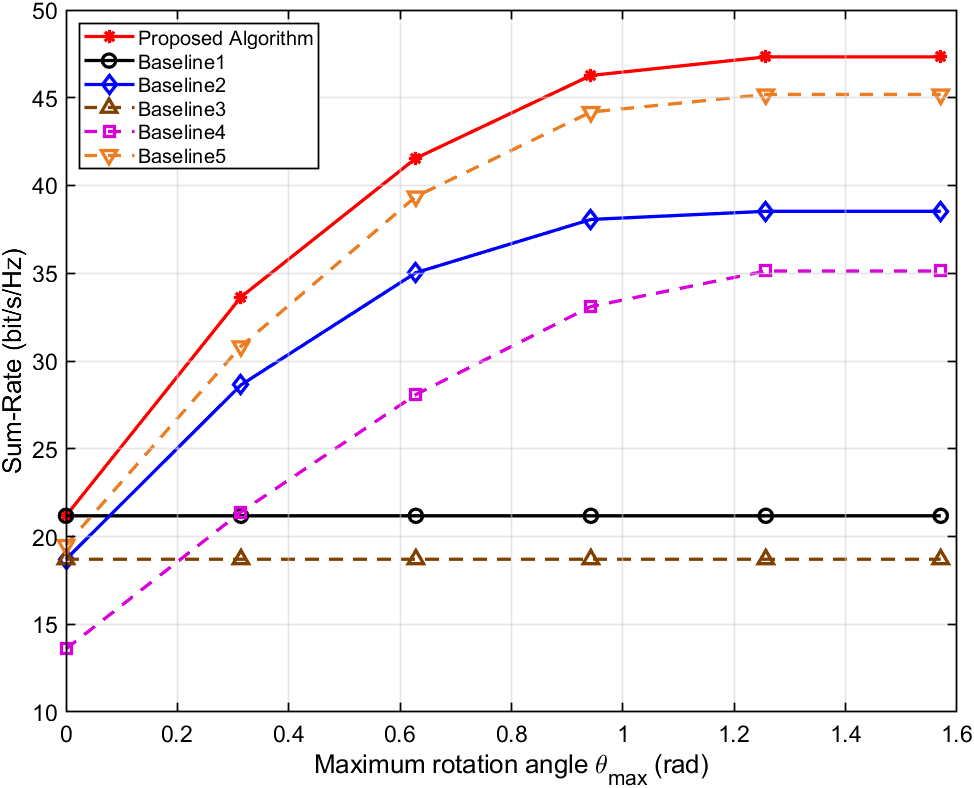}
	\caption{Sum-rate versus the maximum RA rotation angle.}
	\label{fig:simulation_theta}
\end{figure}

Fig.~\ref{fig:simulation_theta} examines the impact of the maximum rotation angle \(\theta_{\max}\) on the system sum-rate. As expected, Baselines~1 and~3 remain invariant with respect to \(\theta_{\max}\) due to the fixed RA orientations. In contrast, the sum-rates of the remaining schemes increase monotonically with \(\theta_{\max}\) and gradually approach plateaus at approximately $1.25$ rad, suggesting that the antennas can fully cover the user distribution range. Among the schemes with optimized RA orientations, the proposed algorithm achieves the highest sum-rate, followed by Baselines~5, 2, and~4. At \(\theta_{\max}=0\), the feasible orientation set is limited to the reference directions. Accordingly, the proposed algorithm coincides with Baseline~1, while Baseline~2 coincides with Baseline~3. The gap between these two pairs reflects the performance gain of optimized user association.

\begin{figure}[!t]
	\centering
	\includegraphics[width=0.75\columnwidth]
	{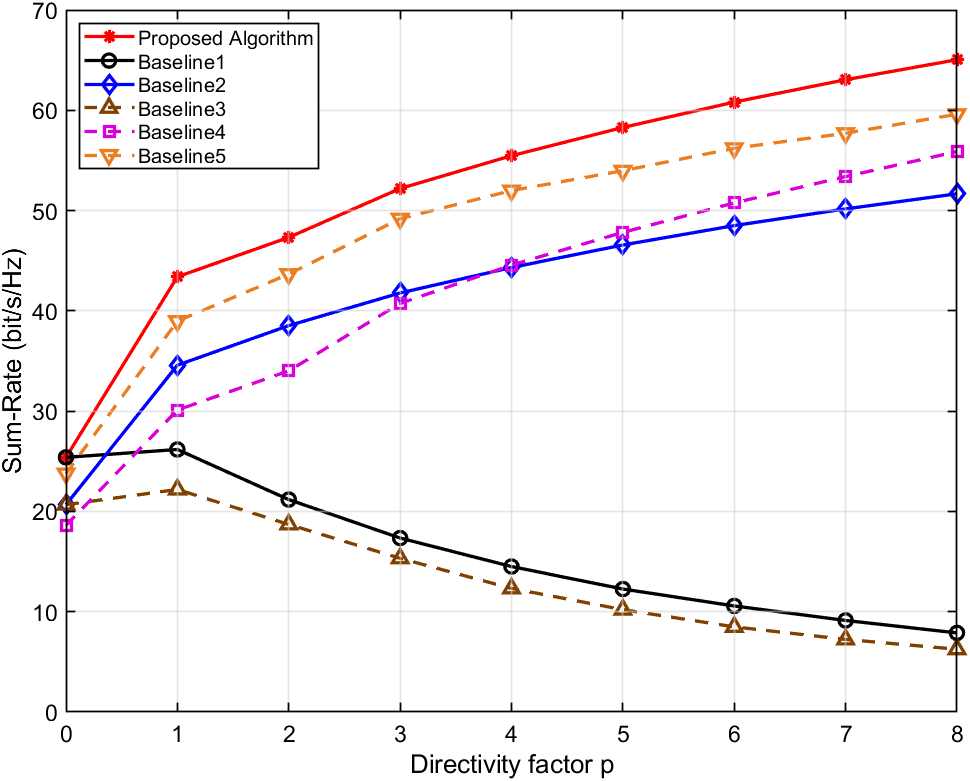}
	\caption{Sum-rate versus the directivity factor.}
	\label{fig:simulation_directivity}
\end{figure}

Fig.~\ref{fig:simulation_directivity} illustrates the effect of the antenna directivity factor \(p\) on the sum-rate. At \(p=0\), the antenna gain in Eq.~\eqref{eq:ra_gain_piecewise} becomes constant and independent of angular misalignment. Hence, optimizing the RA orientations yields no additional benefit. In this case, the proposed algorithm coincides with Baseline~1, and Baseline~2 coincides with Baseline~3. As \(p\) increases, the sum-rates achieved by the proposed algorithm and Baselines~2, ~4, and~5 all increase monotonically, because higher directivity enhances the benefit of steering main lobes toward intended users while suppressing interference. In contrast, the fixed orientation schemes exhibit a slight  improvement at $p=1$, after which the performance declines steadily. The reason is that at low directivity, the increase in \(G_{\max}=2(2p+1)\) outweighs the gain loss caused by boresight misalignment, thereby improving the effective channel gains. However, as \(p\) increases, the narrowing beam pattern aggravates the misalignment loss, which in turn results in the decline.
\begin{figure}[!t]
	\centering
	\includegraphics[width=0.75\columnwidth]
	{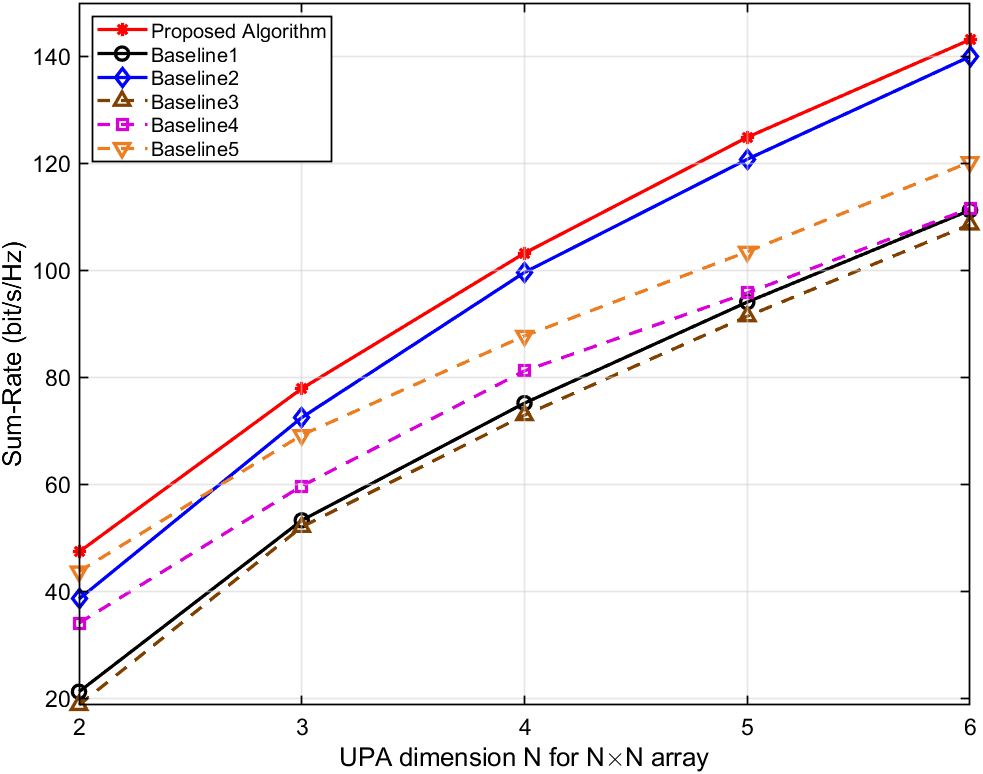}
	\caption{Sum-rate versus the UPA dimension $N$ for an $N\times N$
		array.}
	\label{fig:simulation_array}
\end{figure}

Fig.~\ref{fig:simulation_array} evaluates the effect of the UPA dimension \(N\), where each BS employs an \(N\times N\) array comprising \(M=N^2\) RAs. The sum-rates of all schemes increase monotonically with \(N\). This improvement is attributed to the larger array gain and additional spatial degrees of freedom, which facilitate more effective multi-user beamforming. Furthermore, it is observed that the proposed algorithm consistently achieves the highest sum-rate across all considered array configurations. Meanwhile, its performance gap relative to Baseline~2 narrows as \(N\) increases. This trend indicates that the benefit of optimizing user association becomes less significant for larger arrays. 

\subsection{Performance of the Low-Complexity Algorithms}
\label{subsec:simulation_low_complexity}

In this subsection, we evaluate the low-complexity algorithms developed in Section~\ref{sec:low_complexity_algorithms} against the proposed algorithm in Algorithm~\ref{alg:overall_ao}. All schemes share the same updates for $\mathbf U$, $\mathbf W$, $\mathbf A$, and $\mathbf V$, and differ only in the update of $\mathbf F$. The following low-complexity schemes are considered:
\begin{itemize}
	\item \textbf{Antenna block-based:}
	The RAs are partitioned into $1\times2$ blocks. All RAs within a block share one boresight direction, which is optimized using the Frank-Wolfe update.
	\item \textbf{Discrete candidate-scanning:}
	Each RA retains an independent boresight, but its orientation is selected from a finite candidate set in Eq.~\eqref{eq:discrete_candidate_set}.
	\item \textbf{Combined scheme:}
	Each $1\times2$ block shares one boresight direction, which is updated using discrete candidate scanning.
\end{itemize}

\begin{figure}[!t]
	\centering
	\includegraphics[width=0.75\columnwidth]
	{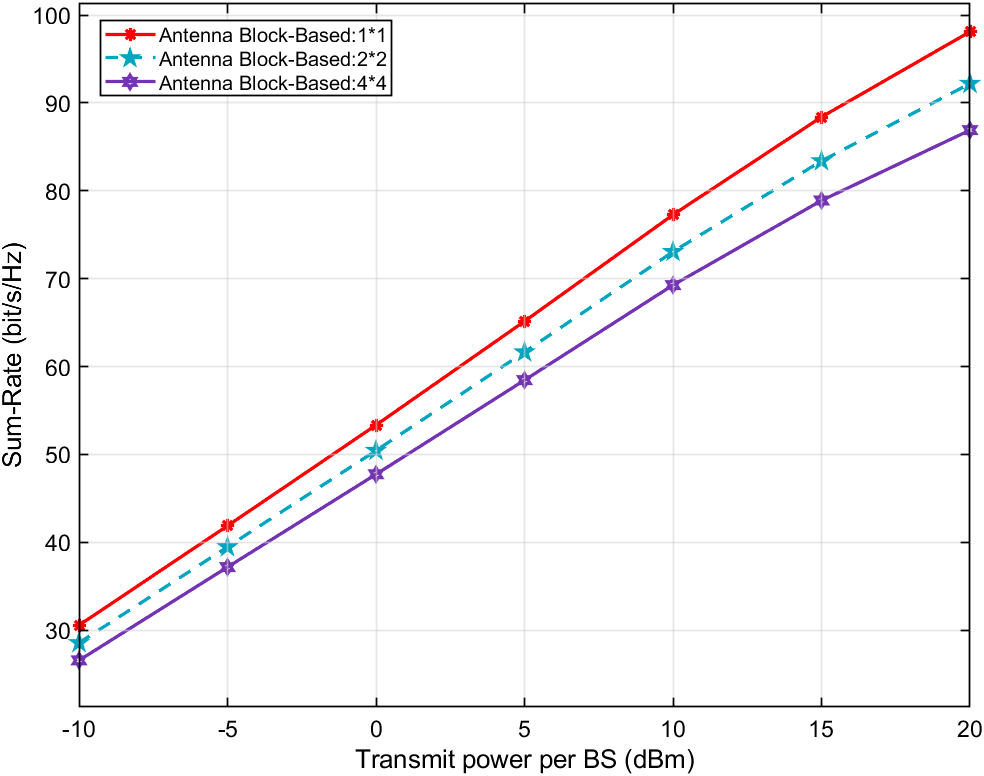}
	\caption{Sum-rate versus transmit power per BS for different antenna block size.}
	\label{fig:simulation_partition}
\end{figure}

Fig.~\ref{fig:simulation_partition} illustrates the effect of antenna block size on the sum-rate for a $4\times4$ UPA. In this comparison, the $1\times1$ curve corresponds to the proposed algorithm in Algorithm~\ref{alg:overall_ao}, where each RA boresight $\mathbf f_{b,m}$ is treated as an independent optimization variable. The $2\times2$ scheme divides the UPA into four blocks, whereas the $4\times4$ scheme assigns a single common orientation to the entire array. The sum-rates of all three schemes increase monotonically with transmit power, and the $1\times1$ scheme consistently achieves the highest sum-rate among them. In particular, the $2\times2$ scheme achieves a sum-rate closer to that of the $1\times1$ scheme, while the $4\times4$ scheme experiences a larger performance degradation. These results indicate that performance loss increases with block size.

\begin{figure}[!t]
	\centering
	\includegraphics[width=0.75\columnwidth]
	{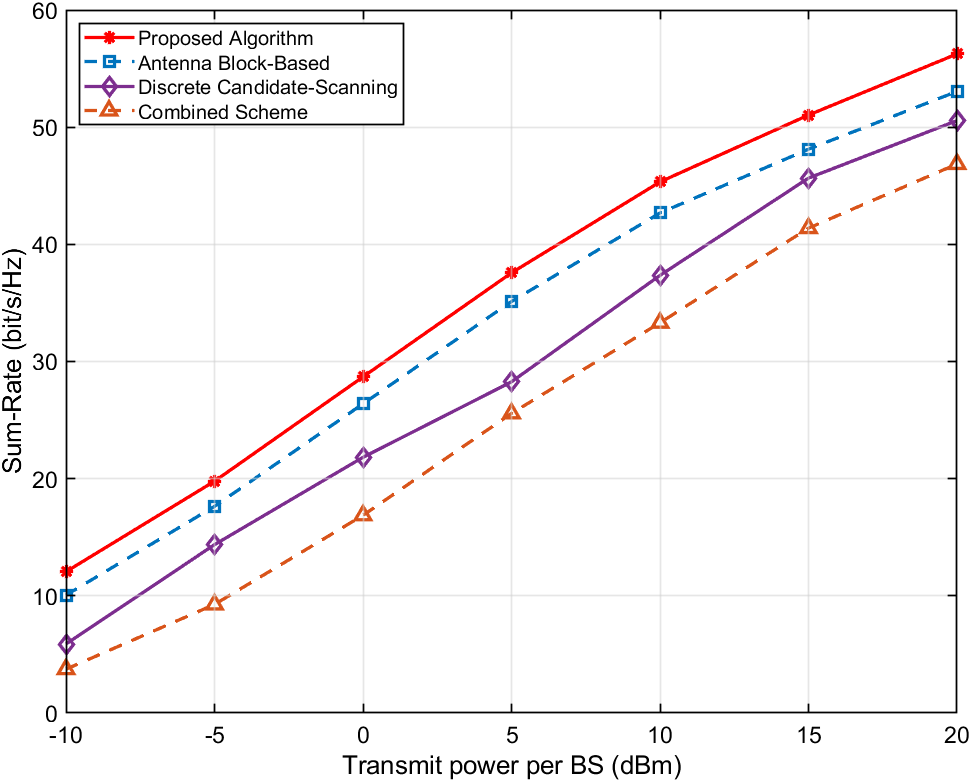}
	\caption{Sum-rate of the low-complexity algorithms versus the transmit
		power per BS.}
	\label{fig:simulation_low_power}
\end{figure}

Fig.~\ref{fig:simulation_low_power} compares the three low-complexity schemes with the proposed algorithm under the $2\times2$ UPA configuration, in which the antenna block-based scheme uses $1\times2$ blocks. The sum-rates of all four schemes increase monotonically with transmit power, and the proposed algorithm consistently achieves the highest sum-rate. The antenna block-based scheme incurs the smallest performance loss relative to the proposed algorithm, because sharing orientations within small blocks retains substantial flexibility in antenna orientation while reducing the number of optimization variables. In contrast, the discrete candidate-scanning scheme suffers a larger degradation, as restricting each boresight to a finite set of user directions sacrifices the flexibility of continuous adjustment.

\section{Conclusion}
\label{sec:conclusions}

In this paper, we have investigated an RA-enabled LAE network in which multiple BSs serve both terrestrial and aerial users. We have formulated a sum-rate maximization problem that jointly optimizes user association, transmit beamforming, and the RA orientations. To address the resulting non-convex problem, we have developed an AO algorithm with projected gradient ascent, the WMMSE algorithm, and the Frank-Wolfe algorithm. We have further developed two low-complexity algorithms that reduce the computational cost of orientation optimization. Simulation results have confirmed that the proposed AO algorithm achieves the highest sum-rate among the evaluated schemes across the examined settings, while the two low-complexity algorithms retain most of the achievable gain with lower orientation update complexity.

\appendices
\section{Derivation of the User-Association Projection}
\label{app:association_projection_derivation}

This appendix derives the Euclidean projection presented in Eq. \eqref{eq:association_projection_update}. For notational convenience, we denote $\mathbf a_k^{(t)}+\lambda^{(t)}\mathbf r_k^{(t)}$ as $\boldsymbol{\xi}_k^{(t)}$. Then the projection of $\boldsymbol{\xi}_k^{(t)}$ onto $\Omega$ is equivalently written as
\begin{subequations}
	\label{prob:appendix_association_projection}
	\begin{align}
		\textup{\textbf P7:}\quad
		\mathbf a_k^{(t+1)}
		=
		\operatorname*{arg\,min}_{\mathbf a\in\mathbb R^{B \times 1}}
		\quad&
		\frac{1}{2}
		\left\|\mathbf a-\boldsymbol{\xi}_k^{(t)}\right\|_2^2
		\label{prob:appendix_association_projection_obj}
		\\
		\mathrm{s.t.}\quad&
		\mathbf a\geq\mathbf 0,
		\label{prob:appendix_association_projection_nonnegative}
		\\
		&
		\mathbf 1^{T}\mathbf a=1.
		\label{prob:appendix_association_projection_sum}
	\end{align}
\end{subequations}
The problem~\textbf P7 is strictly convex and has a unique globally optimal solution. We denote $\boldsymbol\eta_k^{(t)}\geq\mathbf 0$ as the Lagrange multiplier vector associated with constraint \eqref{prob:appendix_association_projection_nonnegative}, and denote $\tau_k^{(t)}\in\mathbb R$ as the Lagrange multiplier associated with constraint~\eqref{prob:appendix_association_projection_sum}, respectively. Then, the Lagrangian is
\begin{equation}
	\begin{aligned}
	\mathcal L_k
	\left(\mathbf a,\tau_k^{(t)},\boldsymbol\eta_k^{(t)}\right)
	={}&
	\frac{1}{2}
	\left\|\mathbf a-\boldsymbol{\xi}_k^{(t)}\right\|_2^2
	\\
	&+
	\tau_k^{(t)}\left(\mathbf 1^{T}\mathbf a-1\right)
	\\
	&-
	\left(\boldsymbol\eta_k^{(t)}\right)^{T}\mathbf a.
	\end{aligned}
	\label{eq:appendix_association_projection_lagrangian}
\end{equation}
The corresponding KKT conditions are
\begin{subequations}
	\label{eq:appendix_association_projection_kkt}
	\begin{align}
		&
		\mathbf a_k^{(t+1)}-\boldsymbol{\xi}_k^{(t)}
		+\tau_k^{(t)}\mathbf 1-\boldsymbol\eta_k^{(t)}
		=\mathbf 0,
		\label{eq:appendix_association_projection_stationarity}
		\\
		&
		\mathbf a_k^{(t+1)}\geq\mathbf 0,
		\label{eq:appendix_association_projection_primal1}
		\\
		&
		\mathbf 1^{T}\mathbf a_k^{(t+1)}=1,
		\label{eq:appendix_association_projection_primal}
		\\
		&
		\boldsymbol\eta_k^{(t)}\geq\mathbf 0,
		\label{eq:appendix_association_projection_complementarity1}
		\\
		&
		\eta_{b,k}^{(t)}a_{b,k}^{(t+1)}=0,
		\quad \forall b\in\mathcal B.
		\label{eq:appendix_association_projection_complementarity}
	\end{align}
\end{subequations}
From Eq. \eqref{eq:appendix_association_projection_stationarity}, the $b$-th component satisfies
\begin{equation}
	a_{b,k}^{(t+1)}
	=
	\xi_{b,k}^{(t)}-\tau_k^{(t)}+\eta_{b,k}^{(t)}.
	\label{eq:appendix_association_projection_component}
\end{equation}
If $a_{b,k}^{(t+1)}>0$, Eq. \eqref{eq:appendix_association_projection_complementarity} requires $\eta_{b,k}^{(t)}=0$, and hence $a_{b,k}^{(t+1)}=\xi_{b,k}^{(t)}-\tau_k^{(t)}$. Otherwise, $a_{b,k}^{(t+1)}=0$. Combining these two cases gives the following closed-form expression for $a_{b,k}^{(t+1)}$:
\begin{equation}
	a_{b,k}^{(t+1)}
	=
	\left[\xi_{b,k}^{(t)}-\tau_k^{(t)}\right]_+,
	\qquad \forall b\in\mathcal B,
	\label{eq:association_projection_closed_form1}
\end{equation}
Substitution Eq. \eqref{eq:association_projection_closed_form1} into Eq.~\eqref{eq:appendix_association_projection_primal} yields
\begin{equation}
	\sum_{b\in\mathcal B}
	\left[\xi_{b,k}^{(t)}-\tau_k^{(t)}\right]_+
	=1.
	\label{eq:association_projection_threshold_condition1}
\end{equation}
Let $\xi_{(1),k}^{(t)}\geq\cdots\geq\xi_{(B),k}^{(t)}$ denote the entries of $\boldsymbol{\xi}_k^{(t)}$ sorted in non-increasing order. The Lagrange multiplier $\tau_k^{(t)}$ is then given by
\begin{equation}
	\tau_k^{(t)}
	=
	\frac{
		\sum_{i=1}^{q_k^{(t)}}\xi_{(i),k}^{(t)}-1
	}{q_k^{(t)}},
	\label{eq:appendix_association_projection_threshold}
\end{equation}
where $q_k^{(t)}$ denotes the number of active components of $\mathbf a_k^{(t+1)}$ and is computed by
\begin{equation}
	\begin{aligned}
	q_k^{(t)}
	={}&
	\max
	\Bigg\{
	j\in\{1,\ldots,B\}:
	\\[-0.5ex]
	&\qquad
		\xi_{(j),k}^{(t)}
	-
		\frac{\sum_{i=1}^{j}\xi_{(i),k}^{(t)}-1}{j}
	>0
	\Bigg\}.
	\end{aligned}
	\label{eq:appendix_association_projection_active_set_size}
\end{equation}
According to the first-order optimality condition,  the optimal $\mathbf a_k^{(t+1)}$ obtained from Eq.~\eqref{eq:association_projection_closed_form1} satisfies
\begin{equation}
	\left(
	\boldsymbol{\xi}_k^{(t)}-\mathbf a_k^{(t+1)}
	\right)^{T}
	\left(
	\mathbf a-\mathbf a_k^{(t+1)}
	\right)
	\leq 0,
	\quad \forall\,\mathbf a\in\Omega.
	\label{eq:appendix_association_projection_variational_inequality}
\end{equation}
Setting $\mathbf a=\mathbf a_k^{(t)}$ and replacing $\boldsymbol{\xi}_k^{(t)}$ with $\mathbf a_k^{(t)}+\lambda^{(t)}\mathbf r_k^{(t)}$ yield
\begin{equation}
	\left(\mathbf r_k^{(t)}\right)^{T}
	\left(\mathbf a_k^{(t+1)}-\mathbf a_k^{(t)}\right)
	\geq
	\frac{1}{\lambda^{(t)}}
	\left\|\mathbf a_k^{(t+1)}-\mathbf a_k^{(t)}\right\|_2^2
	\geq0,
	\label{eq:association_projection_ascent1}
\end{equation}
which means the association update does not decrease the objective.

\section{Derivation of the RA Orientation Gradient}
\label{app:orientation_gradient_derivation}

With $\mathbf U$, $\mathbf W$, $\mathbf A$, and $\mathbf V$ fixed in problem~\textbf P5, only the MSE term depends on $\mathbf F$. Applying the chain rule to the WMMSE objective in Eq. \eqref{prob:wmmse_reformulation} gives
\begin{align}
	\mathbf g_{b,m}
	&=
	\nabla_{\mathbf f_{b,m}}
	\mathcal G_F(\mathbf F)
	\nonumber\\
	&=
	-
	\frac{1}{\ln2}
	\sum_{k\in\mathcal K}
	\sum_{c\in\mathcal B}
	a_{c,k}w_{c,k}
	\nabla_{\mathbf f_{b,m}}e_{c,k}.
	\label{eq:appendix_objective_gradient_initial}
\end{align}
Similarly, $\nabla_{\mathbf f_{b,m}}e_{c,k}$ is given by
\begin{align}
	\nabla_{\mathbf f_{b,m}}e_{c,k}
	&=
	|u_{c,k}|^2
	\nabla_{\mathbf f_{b,m}}T_k
	\nonumber\\
	&\quad-
	2\nabla_{\mathbf f_{b,m}}
	\Re\left\{
	u_{c,k}^{*}\mathbf h_{c,k}^{H}\mathbf v_{c,k}
	\right\},
	\label{eq:appendix_mse_chain_rule}
\end{align}
where $T_k=\sum_{\ell\in\mathcal B}\sum_{j\in\mathcal K}
\left|\mathbf h_{\ell,k}^{H}\mathbf v_{\ell,j}\right|^2+\sigma_k^2$
denotes the total received power at user $k$.
Only channels from BS $b$ depend on $\mathbf f_{b,m}$. Therefore, the received power $T_k$ derivative is
\begin{align}
	\nabla_{\mathbf f_{b,m}}T_k
	&=
	\sum_{j\in\mathcal K}
	\nabla_{\mathbf f_{b,m}}
	\left|
	\mathbf h_{b,k}^{H}\mathbf v_{b,j}
	\right|^2
	\nonumber\\
	&=
	2
	\sum_{j\in\mathcal K}
	\Re\left\{
	\left(
	\mathbf h_{b,k}^{H}\mathbf v_{b,j}
	\right)^{*}
	\times
	\nabla_{\mathbf f_{b,m}}
	\left(
	\mathbf h_{b,k}^{H}\mathbf v_{b,j}
	\right)
	\right\}.
	\label{eq:appendix_channel_power_gradient}
\end{align}
Similarly, 
\begin{align}
	&\nabla_{\mathbf f_{b,m}}
	\Re\left\{u_{c,k}^{*}\mathbf h_{c,k}^{H}\mathbf v_{c,k}\right\}=
	\begin{cases}
		\displaystyle
		\Re\left\{
		u_{b,k}^{*}
		\nabla_{\mathbf f_{b,m}}
		\left(
		\mathbf h_{b,k}^{H}\mathbf v_{b,k}
		\right)
		\right\},
		& b=c,\\[8pt]
        \mathbf 0,
		& b\neq c.
	\end{cases}
	\label{eq:appendix_desired_term_gradient}
\end{align}
For notational convenience, let $\boldsymbol\psi_{b,k,j,m}$ denote the gradient of the effective channel coefficient $\mathbf h_{b,k}^{H}\mathbf v_{b,j}$ with respect to the orientation $\mathbf f_{b,m}$. By Eq. \eqref{eq:exact_channel_vector}, $\mathbf f_{b,m}$ affects only the $m$-th entry of $\mathbf h_{b,k}$.
Therefore,
\begin{align}
	\boldsymbol\psi_{b,k,j,m}
	&=
	\nabla_{\mathbf f_{b,m}}
	\left(
	\mathbf h_{b,k}^{H}\mathbf v_{b,j}
	\right)
	\nonumber\\
	&=
	v_{b,j,m}
	\left(
	\nabla_{\mathbf f_{b,m}}h_{b,k,m}
	\right)^{*}.
	\label{eq:appendix_effective_channel_derivative}
\end{align}
Differentiating Eq. \eqref{eq:element_channel_exact} with respect to $\mathbf f_{b,m}$ yields
\begin{equation}
	\begin{aligned}
	\nabla_{\mathbf f_{b,m}}h_{b,k,m}
	&=
	\frac{p\sqrt{\beta_0G_{\max}}}{r_{b,k,m}}
	\exp\left(
	-\mathrm j\frac{2\pi}{\lambda}
	\boldsymbol{\ell}_{b,k}^{T}\mathbf r_{b,m}
	\right)
	\\
	&\quad\times
	\left[
	\mathbf f_{b,m}^{T}\mathbf u_{b,k,m}
	\right]_{+}^{p-1}
	\mathbf u_{b,k,m}.
	\end{aligned}
	\label{eq:appendix_element_channel_gradient_general}
\end{equation}
We adopt the convention that $[x]_{+}^{p-1}=0$ whenever $[x]_{+}=0$, which selects the zero Clarke subgradient at the visibility boundary \cite{peng2026rotatable}. Substituting Eqs. \eqref{eq:appendix_channel_power_gradient}--\eqref{eq:appendix_effective_channel_derivative} into Eq. \eqref{eq:appendix_mse_chain_rule}, and then substituting the result into \eqref{eq:appendix_objective_gradient_initial}, yields
\begin{equation}
	\begin{aligned}
	\mathbf g_{b,m}
	&=
	\nabla_{\mathbf f_{b,m}}\mathcal G_F
	\\
	&=
	\frac{2}{\ln 2}
	\sum_{k\in\mathcal K}
	\Bigg[
	a_{b,k}w_{b,k}
	\Re\!\left\{
	u_{b,k}^{*}
	\boldsymbol\psi_{b,k,k,m}
	\right\}
	\\
	&-
	\left(
	\sum_{c\in\mathcal B}
	a_{c,k}w_{c,k}|u_{c,k}|^2
	\right)
	\\
	&\quad
	\times
	\sum_{j\in\mathcal K}
	\Re\!\left\{
	\left(
	\mathbf h_{b,k}^{H}\mathbf v_{b,j}
	\right)^{*}
	\boldsymbol\psi_{b,k,j,m}
	\right\}
	\Bigg].
	\end{aligned}
	\label{eq:appendix_orientation_gradient_expanded}
\end{equation}


\bibliographystyle{IEEEtran}
\bibliography{reference}

\end{document}